\documentclass[apj]{emulateapj}
\usepackage{epsfig}
\usepackage{subfigure}

\def\Aa{-2.5038e-3}
\def\Ab{0.12912}
\def\Ac{-2.4273}
\def\Ad{19.980}
\def\Ae{-60.931}

\def\Ba{3.0668e-3}
\def\Bb{-0.15902}
\def\Bc{3.0365}
\def\Bd{-25.320}
\def\Be{82.605}

\def\Ca{-1.5465e-5}
\def\Cb{7.5396e-4}
\def\Cc{-1.2836e-2}
\def\Cd{9.6434e-2}
\def\Ce{-0.27166}

\def\Da{2.7978e-7}
\def\Db{-1.5572e-5}
\def\Dc{3.1957e-4}
\def\Dd{-2.8543e-3}
\def\De{9.3191e-3}

\def\Ea{-6.4215e-6}
\def\Eb{3.5358e-4}
\def\Ec{-7.1463e-3}
\def\Ed{6.2522e-2}
\def\Ee{-0.19743}

\def\Pr{{\rm Pr}}
\def\pipl{\pi_{\rm pl}}
\def\piBB{\pi_{\rm BB}}

\def\piHT{\pi_{\rm HT}}

\def\Kepler{\textit{Kepler~}}

\begin{document}

\title{On The Low False Positive Probabilities of \textit{KEPLER} Planet Candidates}

\author{Timothy D. Morton\altaffilmark{1},John Asher Johnson\altaffilmark{1,2}}

\email{tdm@astro.caltech.edu}
\email{johnjohn@astro.caltech.edu}

\altaffiltext{1}{Department of Astrophysics,
  California Institute of Technology, MC 249-17, Pasadena, CA 91125}
\altaffiltext{2}{NASA Exoplanet Science Institute (NExScI)}

\begin{abstract}
We present a framework to conservatively estimate the probability that any particular planet-like transit signal observed by the \Kepler mission is in fact a planet, prior to any ground-based follow-up efforts.  We use Monte Carlo methods based on stellar population synthesis and Galactic structure models, and report \textit{a priori} false positive probabilities for every Kepler Object of Interest in tabular form, assuming a 20\% intrinsic occurrence rate of close-in planets in the radius range $0.5 R_\oplus < R_p < 20~R_{\rm \oplus}$.   Almost every candidate has FPP $<10\%$, and over half have FPP $<5\%$.  This probability varies most strongly with the magnitude and Galactic latitude of the \Kepler target star, and more weakly with transit depth.  We establish that a single deep high-resolution image will be an extremely effective follow-up tool for the shallowest (Earth-sized) transits, providing the quickest route towards probabilistically ``validating'' the smallest candidates by potentially decreasing the false positive probability of an earth-sized transit around a faint star from $>$10\% to $<$1\%.   On the other hand, we show that the most useful follow-up observations for moderate-depth (super-Earth and Neptune-sized) candidates are shallower AO imaging and high S/N spectroscopy. Since \Kepler has detected many more planetary signals than can be positively confirmed with ground-based follow-up efforts in the near term, these calculations will be crucial to using the ensemble of \Kepler data to determine population characteristics of planetary systems.   We also describe how our analysis complements the \Kepler team's more detailed BLENDER false positive analysis for  planet validation.

\end{abstract}

\section{Introduction}

In the wake of the first full release of planet candidates from the \Kepler mission \citep{koch98,borucki08,borucki11}, the study of the properties of exoplanetary systems has entered a new era.  For the first time there exists a large uniform sample of transiting planets largely unaffected by the detection challenges and selection effects inherent in ground-based searches \citep[e.g.]{gaudi05,gaudietal05}, enabling the first clear glimpse of the population of exoplanets down to the size of Earth as well as the first opportunity to study planet radii at large orbital separations.  However, follow-up observations to unambiguously confirm individual signals are time-consuming and difficult (or impossible), especially for fainter stars and smaller planets.  Consequently, in order to understand what the population of \Kepler transit-like signals can tell us about the population of exoplanets in general, the problem of astrophysical false positives must be understood.

From the early days of planet transit searches, eclipsing binary systems masquerading as transit signals have plagued detection efforts \citep{OGLEFPs,odonovan06,XOFPs,almenara09}.  Generally speaking, there are three types of astrophysical false positive: a grazing eclipsing binary, a dwarf star eclipsing a giant star, and a blended eclipsing binary system, which may be either a hierarchical triple system or an unassociated binary blended within the aperture of a target star \citep{torres04}.\footnote{In this paper we consider ``false positives'' to be purely stellar configurations mimicking transiting planet signals.  For discussion of scenarios involving ``blended planets,'' see Appendix.} 

The remarkable photometric precision that \Kepler is delivering \citep[$\sim$30 ppm]{jenkins10b} allows for an immediate simplification of the false positive landscape.  \citet{batalha10b} explain the multitude of ways that certain common false positive scenarios can be identified from \Kepler photometry alone.  For example, grazing eclipsing binaries can be identified by their V-shaped transits, and the giant-eclipsed-by-a-dwarf scenario can be avoided both by the comprehensive work that went into assembling the Kepler Input Catalog \citep{latham05,batalha10a} and by the ability to photometrically identify giants by their elevated levels of stellar variability compared with dwarf stars \citep{basri10}. 
Even many blended binaries can be identified from the \Kepler photometry and astrometry alone, by looking for a shift in the center of light, e.g~the ``rain diagrams'' of \citet{jenkins10}.  However, some blended binary scenarios remain undetectable by this technique, especially those in hierarchical triple systems, and so a detailed understanding of the false positive problem for \Kepler requires a detailed understanding of the probability of encountering such blend scenarios.  

The \Kepler team has proven that extremely careful and detailed analyses of individual systems can ``validate'' planets probabilistically by combining various follow-up observations with modeling the light curves of all possible false positive scenarios with the so-called BLENDER software \citep{torres11}.  However, this method is computationally expensive and labor-intensive, rendering it a time-consuming process, and only three BLENDER-validated planets having been revealed to date: Kepler-9d \citep{torres11}, Kepler-11f \citep{lissauer11}, and Kepler-10c \citep{fressin11}.  With dedicated supercomputer resources coming online for the \Kepler team's use, this number will certainly rise, but the fact remains that it will be a long time before the BLENDER method can be applied to any large number of the \Kepler candidates (\Kepler team, 2011, private comm.); in the meantime statistical interpretations of the candidate sample will rely on statistical assumptions of the false positive rate.


There has been significant previous effort in the literature dedicated to predicting the expected rate of false positive transit signals.  \citet{brown03} pioneered this work by predicting the rates of different types of false positives and Jovian planet detections for a variety of different surveys, including the then-future \Kepler mission.  \citet{evans10} greatly extend this work by deriving detection and false positive rates from full-scale bottom-up simulations of synthetic ground-based transit surveys, taking into account all false positive possibilities and many details not included by \citet{brown03}.  We continue in the tradition of these authors with an analysis directly applicable to the \Kepler mission, approaching from a slightly different angle.  
Instead of focusing on predicting an overall number or  expected rate of planet detections or false positives, we instead seek a simple answer 
to the following question:  ``What is a conservative estimate of the probability that an observed apparent transit signal is in fact a true transiting planet?''  By framing the issue in this manner we are able to sidestep the complex issue of detectability, as our analysis assumes a transit-like signal has been detected.

Our philosophy in this work is not to take into account all conceivable details of transit signals, but rather to consider only those which are most salient: the brightness of the \Kepler target star, its location in the field, and transit signal depth.
The details we choose not to address in this work (notably transit period and duration) are those we judge would add uncertainty to our calculations while tending to only decrease our estimates of the false positive probability.  Thus we are able to keep our analysis straightforward, yet remain confident we are calculating conservative upper limits to the probability that any given \Kepler transit signal might be a false positive.  As we show in \S\ref{framework} and again in \S\ref{detailed}, even these conservative upper limits are enough to indicate that \Kepler planet candidates will only rarely turn out to be false positives. 


\section{Basic Bayesian Framework}
\label{framework}
The probability that a given transit signal is of planetary origin may be expressed as the following, according to Bayes' theorem:
\begin{equation}
\label{eq:simplebayes}
\Pr({\rm planet} ~|~{\rm signal}) = \frac{\Pr({\rm signal~|~planet}) \Pr({\rm planet})}{\Pr({\rm signal})}.
\end{equation}
In this framework $\Pr({\rm signal~|~planet})$ is the probability of obtaining the observed signal given that there is a transiting planet on an orbit of a particular period.  This factor is known as the \textit{likelihood} of the signal under the planet hypothesis, and we will abbreviate it as $\mathcal L_{\rm pl}$.  $\Pr({\rm planet})$ is the probability of a star hosting a transiting planet (the occurrence rate of planets times the transit probability), which must enter the calculation as an \textit{a priori} assumption.  Thus we call this factor, according to Bayesian convention, the \textit{prior} on planets, and designate it $\pi_{\rm pl}$.

Since there are only two possible origins of a transit-like signal (planet or false positive), the denominator of Equation \ref{eq:simplebayes} can be rewritten as marginalizing over the possible models:
\begin{equation}
\label{eq:psignal}
\Pr({\rm signal}) =  \mathcal L_{\rm pl} \pi_{\rm pl} + \mathcal L_{\rm FP}  \pi_{\rm FP}.
\end{equation}
Using our convention, $\mathcal L_{\rm FP}$ and $\pi_{\rm FP}$ are the likelihoods and priors for a false positive signal.
The false positive term can be further broken down accounting for the two specific false positive scenarios we are exploring: the blended eclipsing binary (BB) and the hierarchical eclipsing triple (HT), allowing Equation \ref{eq:simplebayes} to be rewritten as the following:
\begin{equation}
\label{eq:simplebayesLs}
\Pr({\rm planet} ~|~{\rm signal}) = \frac{\mathcal L_{\rm pl} \pi_{\rm pl} } { \mathcal L_{\rm pl} \pi_{\rm pl} + 
\mathcal L_{\rm BB} \pi_{\rm BB} + \mathcal L_{\rm HT} \pi_{\rm HT} }.
\end{equation}

In general, the likelihoods depend on the particularities of the transit signal and enable discrimination between models depending on the transit depth, shape, or period.  For now we ignore these details, assuming for the moment that we have no knowledge of the differences between the kind of transit signals to expect from planets and from false positives.  This enables us to write a simplified version of Eq.~\ref{eq:simplebayesLs}:
\begin{equation}
\label{eq:simplebayespriorsonly}
\Pr({\rm planet} ~|~{\rm signal}) \approx \frac{\pi_{\rm pl} } {\pi_{\rm pl} + \pi_{\rm BB} + \pi_{\rm HT} }.
\end{equation}
We then define the ``false positive probability'' (FPP) as the complement of this probability:
\begin{equation}
\label{eq:FPP}
{\rm FPP} = 1 - \Pr({\rm planet} ~|~{\rm signal})
\end{equation}
Thus, before considering any detailed information of a particular light curve, the probability that an observed transit signal is actually a false positive depends only on the relative occurrence rates of planets and the false positive scenarios.  As mentioned above, $\pipl$ is simply an assumed occurrence rate of planets times the transit probability; we explain how we determine $\piBB$ and $\piHT$ in the following subsections.  We explain first this priors-only framework in order to elucidate what dominates our final results, but in \S\ref{detailed} we will include the likelihoods we removed in Equation \ref{eq:simplebayespriorsonly}, taking into account dependence on the depth of the transit signal. 



\subsection{Blended Binaries}
\label{piBB}

The probability of a transit-mimicking binary system to be blended within the aperture of a \Kepler target star ($\piBB$) can be broken down into the following way:

\begin{equation}
\label{eq:piBB}
\piBB = \Pr({\rm blend}) \cdot \Pr({\rm appropriate~eclipsing~binary}).
\end{equation}
The first factor here is the probability for a potentially blending star to be projected within a given radius of a \Kepler star, and the second is the probability for that star to be an eclipsing binary system that can appropriately mimic a planetary transit.  

To calculate these probabilities, we use the stellar population synthesis and Galactic structure code TRILEGAL (TRIdimensional modeL of thE GALaxy; \citet{girardi05}), which is publicly available on the web\footnote{http://stev.oapd.inaf.it/cgi-bin/trilegal}.  TRILEGAL simulates the physical and photometric properties of the stars along a given line of sight, using various stellar evolution grids \citep{girardi02,chabrier00} and a Galactic model that includes a halo, thin and thick disks, and a bulge.  All of our simulations use a Chabrier lognormal IMF \citep{chabrier01} and default TRILEGAL values for the Galactic structure parameters, including a squared hyperbolic secant structure for the thin disk, an exponential structure for the thick disk, and an oblate spheroid for the halo.  

\subsubsection{Probability of a blend}
\label{pblend}

\begin{figure}[!t]
	\includegraphics[width=3.8in]{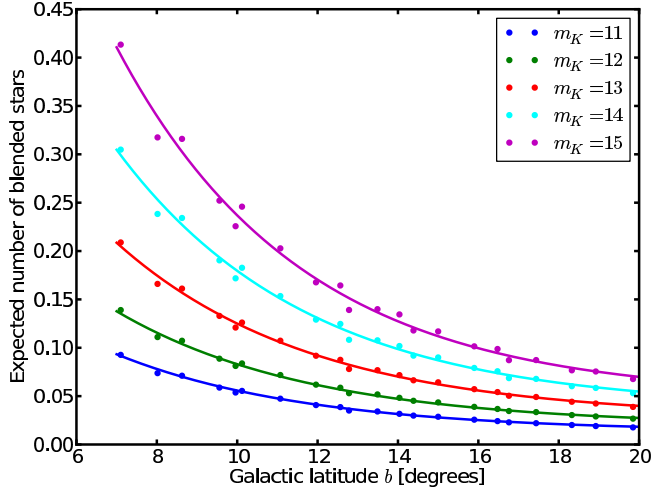}
	\caption{The probability for a possibly blending star to be projected within 2$''$ of a \Kepler target star, as a function of Galactic latitude, as determined by TRILEGAL simulations.  The plotted points are simulations; the lines are the exponential fits as described in Equation \ref{eq:pblendfn}.}
	\label{fig:pblendofb}
\end{figure}

The blend probability can be calculated by determining the average sky density (e.g.~stars per square arcsec) of stars faint enough so as not to be obviously present 
yet bright enough to possibly mimic a transit.  The first condition is somewhat subjective, and we conservatively say that a star must be more than 1 magnitude fainter than the \Kepler primary in order to be able to hide undetected within the \Kepler aperture.  In practice the true value is probably significantly fainter, but this approximation will lead to only a small overestimate of the blended star probability, as there are many more faint than bright stars.  

The faint condition can be determined by noting that in order for a blended eclipsing binary system to mimic a transit of fractional depth $\delta$, the blended system must comprise more than a fraction $\delta$ of the total flux within the \Kepler aperture.  This condition may be expressed as the following:
 \begin{equation}
 \label{eq:blender}
m_{K,{\rm bin}} -  m_{K,{\rm target}} = \Delta m_K = -2.5 \log_{10}(\delta),
\end{equation}
where $m_{K,{\rm bin}}$ is the total apparent Kepler magnitude of the blended binary system and $m_{K,{\rm target}}$ is the magnitude of the \Kepler target star.  A transit depth of $\delta = 0.01$ corresponds to $\Delta m_K = 5$; for $\delta = 10^{-3}$, $\Delta  m_K = 7.25$; and for $\delta = 10^{-4}$ (approximately an Earth-sized transit of a Solar-radius star), $\Delta m_K = 10$.  This means that no binary system fainter than $m_K = 24$ can possibly mimic a $\delta = 10^{-4}$ transit around a  $m_K = 14$ star, which is a typical magnitude for a \Kepler target.

Using TRILEGAL, we determine the sky density of stars in this magnitude range within in the \Kepler field, and thus the probability of one by chance being projected close to a \Kepler target star, by simulating a 10 deg$^2$ field centered on the center of the \Kepler field.   We then simply count the stars within the desired range of Kepler magnitude (which TRILEGAL provides).  As a fiducial example, the average density of stars between $m_K = 15$ and $m_K = 23.25$, the range corresponding to a $\delta = 10^{-4}$ transit of a $m_K = 14$ star, is 0.0085 stars$\cdot$arcsec$^{-2}$.  The probability of any given small circle on the sky containing one of these stars is then simply the area of the circle multiplied by this density.  Continuing this example, ($m_K = 14,\delta=10^{-4}$) the probability of such a star being within 2$''$ of a \Kepler target star is 0.11.

\begin{figure}[!t]
	\includegraphics[width=3.8in]{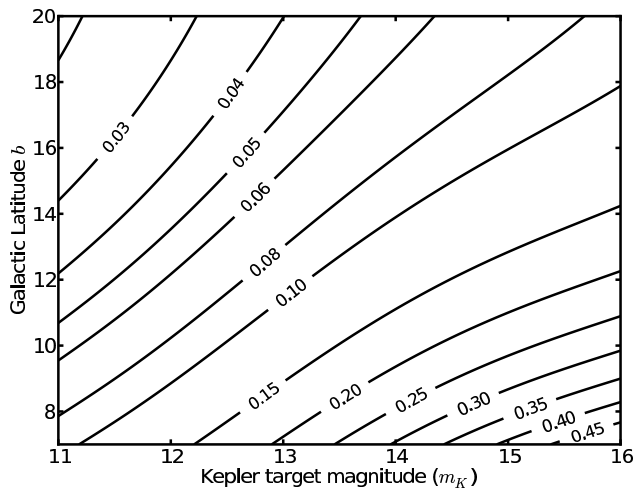}
	\caption{The probability for a possibly blending star to be projected within 2$''$ of a \Kepler target star, as a function of both Galactic latitude, and target star magnitude, as determined by TRILEGAL simulations.}
	\label{fig:pblendbasic}
\end{figure}

However, because the \Kepler field is quite extended and centered only a few degrees off the Galactic plane, there is a considerable gradient in background stellar density across the field that must be accounted for.  To accomplish this, we simulate 21 different 5 deg$^2$ fields, each centered on one of the \Kepler double-CCD squares.  The resulting probabilities are plotted in Figure \ref{fig:pblendofb} as a function of Galactic latitude, for the magnitude ranges corresponding to $m_K = 11,12,13,14,$ and 15.  Recognizing that this blend probability appears to be exponentially related to Galactic latitude $b$, and that the nature of the exponential depends on $m_K$, we fit an analytic expression of the following form:
\begin{equation}
\label{eq:pblendfn}
p_{\rm blend}(b,m_K) = C(m_K) + A(m_K) e^{-b / B(m_K)},
\end{equation}
where $A$, $B$, and $C$ are all polynomial functions of Kepler magnitude, with the coefficients listed in Table \ref{polytable}.  These fits are valid between $m_K$ values of 11 and 15, and $b$ values between 7$^\circ$ and 20$^\circ$ (the approximate extent of the \Kepler field).  Figure \ref{fig:pblendbasic} graphically illustrates the behavior of Equation \ref{eq:pblendfn}.

\begin{deluxetable}{cccccc}
	\tabletypesize{\footnotesize}
	\tablecolumns{6}
	\tablecaption{Polynomial coefficients\tablenotemark{1} for Equation \ref{eq:pibbfn}}
	\tablehead{ \colhead{} &\colhead{$c_0$} & \colhead{$c_1$} &  \colhead{$c_2$} &  \colhead{$c_3$} & \colhead{$c_4$}}
	\startdata
	$A$ &\Aa&\Ab&\Ac&\Ad&\Ae\\
	$B$ &\Ba&\Bb&\Bc&\Bd&\Be\\
	$C$ &\Ca&\Cb&\Cc&\Cd&\Ce\\
	$D$ &\Da&\Db&\Dc&\Dd&\De\\
	$E$ &\Ea&\Eb&\Ec&\Ed&\Ee
	\enddata
	\tablenotetext{1}{This table lists the polynomial coefficients for the empirical fits to how the blended binary false positive probability as a function of Galactic latitude changes with Kepler magnitude $m_K$.  $A,B,C,D$, and $E$ are functions of $m_K$, valid between $m_K=11$ and $m_K=16$.  The polynomials are of the form $c_0 + c_1m_K + c_2m_K^2 + c_3m_K^3 + c_4m_K^4$.}
	\label{polytable}
\end{deluxetable}

\begin{deluxetable*}{ccccccc|c}
	\tabletypesize{\footnotesize}
	\tablecolumns{8}
	\tablecaption{Predicted False Positive Probabilities: Basic Framework}
	\tablehead{ \colhead{Experiment} &\colhead{Threshold} & \colhead{Blend Radius} &  \colhead{$<\# {\rm ~blends}>$} &   \colhead{$\piBB$} &   \colhead{$\piHT$} &  \colhead{$\pipl$} &  \colhead{FPP} \\ (1) & (2) & (3) & (4) & (5) & (6) & (7) & (8)}
	\startdata
	\Kepler & $10^{-4}$ & $2''$ & 0.11 &   $1.1\times 10^{-4}$ &  $1.2 \times 10^{-4}$ &  0.01 &  \textbf{0.02} \\
	Wide-Field Survey (e.g.~HATNet) & 0.005 & $14''$ & 1.67 & 0.0014 & $2.7\times 10^{-4}$ & $5 \times 10^{-4}$ & \textbf{0.77} \\
	CoRoT & $10^{-3}$ & $10''$ & 2.71 &  0.0035 &  $1.4 \times 10^{-4}$ &  0.01 &  \textbf{0.27} \\
	\enddata
	\tablenotetext{(1)}{Name of a transit survey}
	\tablenotetext{(2)}{Fractional depth detection threshold}
	\tablenotetext{(3)}{Effective aperture size inside which a blended star might reside.  \Kepler can restrict this radius to 2$''$ by centroid analysis.}
	\tablenotetext{(4)}{The expected number of blending stars expected per aperture, based on estimates of the density of stars within the possibly-blending magnitude range for each experiment.}
	\tablenotetext{(5)}{The rate we calculate for the blended eclipsing binary false positive scenario}
	\tablenotetext{(6)}{The rate we calculate for the hierarchical eclipsing triple false positive scenario}
	\tablenotetext{(7)}{The assumed rate of detectable transiting planets}
	\tablenotetext{(8)}{False positive probability $= \pipl/(\piBB + \piHT + \pipl)$}
	\label{resultstable}
\end{deluxetable*}

\subsubsection{Probability of an appropriate eclipsing binary}
\label{pappropriate}

The probability that a blended star is an appropriately configured eclipsing binary system depends first on the binary fraction of blending stars, and secondly on both the distribution of binary properities and the magnitude of the \Kepler target star.  Of central importance is that in order for a blended binary to successfully mimic a \Kepler planet transit candidate, it must both have a diluted primary eclipse shallow enough to look like a planet and a diluted secondary eclipse shallow enough so as not to be detected.  

The apparent fractional ``transit'' depth of a blended binary system depends on the intrinsic binary system eclipse depth $\delta_b$, and the relative apparent magnitudes of the \Kepler target star and the blended system:
\begin{equation}
\label{eq:blenddepth}
\delta = \delta_b \cdot 10^{-0.4(m_{K,{\rm bin}} - m_{K,{\rm target}}) }.
\end{equation}

The primary and secondary eclipse depths of the binary system are the following:
\begin{equation}
\label{eq:pridepth}
\delta_{b,{\rm pri}} = \displaystyle \frac{\left(\frac{R_2}{R_1}\right)^2 F_1}{F_1+F_2},
\end{equation}
and
\begin{equation}
\delta_{b,{\rm sec}} = \frac{F_2}{F_1 + F_2},
\label{eq:secdepth}
\end{equation}
where $R_1$ and $F_1$ are the stellar radius and flux in the \Kepler band of the larger of the two stars, and $R_2$ and $F_2$ are of the smaller star.

The conditions we define for a binary to be ``appropriate'' are for the diluted primary eclipse depth to be between $0.02$ and $10^{-4}$ (shallow enough to look like a planet, but still detectable), and for the diluted secondary to be shallower than $10^{-4}$ (undetectable).  We recognize that ``detectability'' of a transit is a function of more than just the transit depth, but for our purposes we use a depth of $10^{-4}$ as the detection threshold.  A more detailed population study based on \Kepler candidates should use rather the signal-to-noise ratio of a transit as the criteron for detectability \citep{beatty08}.  However, as our framework deals with how to interpret signals once they are detected, careful detectability analysis in unnecessary.

To calculate the probability of all these conditions being met (a star being binary and being ``appropriate''), we use the TRILEGAL simulations and assume binary properties according to the work of \citet{raghavan}.   That is, we assume a flat mass ratio distribution between 0.1 and 1 (\citet{raghavan} actually observes the distribution to be flat between about 0.2 and 1, but we extend it to 0.1 to be more conservative).

For each star in a particular TRILEGAL line-of-sight simulation that lies in the appropriate magnitude range (\S\ref{pblend}), we first randomly assign it to be a binary or not and then calculate what the primary and secondary diluted depths would be if the system were eclipsing and blended with a \Kepler target star of a particular magnitude.  
$R_1$ and $F_1$ are provided by TRILEGAL\footnote{This properly accounts for the possibility that the blend might be an evolved system; e.g.~a dwarf star eclipsing a giant.}, and we determine $R_2$ and $F_2$ based on a randomly assigned mass ratio and the Padova models at the age of the primary.
Given these system parameters, we can then randomly determine if each system undergoes a non-grazing eclipse, according to the probability that each system will be in such an orientation:
\begin{equation}
\label{eq:peclipse}
\Pr({\rm eclipse}) = \frac{R_1-R_2}{a},
\end{equation}
where $a$ is the orbital semi-major axis, determined from Kepler's law.
  
From this procedure, using a \Kepler target star of $m_K=14$, an orbital period of 10 days, and a line-of-sight simulation at the center of the \Kepler field, we find that  1.25\% of binaries have non-grazing eclipses and about 20\% of those eclipsing binaries are ``appropriate.''  Combined with a $\sim$40\% binary fraction\footnote{To be precise, we actually use a binary fraction function that increases with stellar mass: 40\% for $M < M_\odot$, 50\% for $M_\odot < M < 1.5 M_\odot$, and 75\% for $M>1.5 M_\odot$, roughly adapted from Figure 12 in \citet{raghavan}.  This is a conservative estimate of the binary fraction, as the Raghavan Figure includes multiple systems as well as binaries.} , this results in a probability of 0.001 for a star to be an appropriate eclipsing binary, giving a value of $\piBB = 0.11 \times 0.001 = 1.1 \times 10^{-4}$ for the center of the \Kepler field.

As in \S\ref{pblend}, we empirically investigate how this probability changes as a function of galactic latitude and target star magnitude.  We find the behavior for any particular magnitude is well described by a shallow linear relation in $b$:
\begin{equation}
\label{eq:pecbfn}
\Pr({\rm appropriate~ecl.~binary}) = bD(m_K) + E(m_K),
\end{equation}
where again the variation of the values of the coefficients $D$ and $E$ is modeled well with a polynomial in $m_K$ (Table \ref{polytable}).
 
Multiplying Equation \ref{eq:pecbfn} with Equation \ref{eq:pblendfn} then gives a full analytic expression for the probability of a star of given Kepler magnitude at a given Galactic latitude to be blended with an eclipsing binary system able to mimic a planetary transit:
\begin{eqnarray}
\label{eq:pibbfn}
\piBB(m_K,b) = & \left[ C(m_K) + A(m_K) e^{-b/B(m_K)} \right] \times \nonumber \\
& \left[ b D(m_K) + E(m_K) \right],
\end{eqnarray}
where $A,B,C,D,$ and $E$ are polynomial functions of $m_K$ with coefficients given in Table \ref{polytable}.  


\subsection{Hierarchical Triples}

\label{piHT}
The probability that a \Kepler target star is in fact a hierarchical triple system configured such that it might be able to mimic a planetary transit ($\piHT$) can be broken down as follows:
\begin{equation}
\label{eq:piHT}
\piHT = \Pr({\rm triple}) \cdot \Pr({\rm eclipsing~and~appropriate}).
\end{equation}
The first factor is simply the frequency of triple systems, which \citet{raghavan} determine to be 8\% for sun-like stars.  The fraction of triple systems that are of appropriate configuration can be determined by using the same conditions as we used above in \S\ref{pappropriate}.  That is, we require the diluted eclipse depths (Eqs. \ref{eq:blenddepth}-\ref{eq:secdepth}) to be between $0.02$ and $10^{-4}$, except this time one of the three triple components provides the diluting flux.

We assume two different hierarchical possibilities for triple systems.  Referring to the three components in order of descending mass as A, B, and C, the triple system may either be set up as A + BC, where B \& C are the closer potentially eclipsing pair and A is the diluting star, or as AC + B, with A \& C as the closer pair and B diluting.  We ignore the case AB + C because the faintest component being the diluting star would be unable to mimic a planet transit.  

We calculate the probability that a triple system will be eclipsing and ``appropriate''  (again assuming a 10-day orbit) as follows:
\begin{equation}
\label{eq:patriples}
p_a = \int\int A(M_A,q_1,q_2) \Phi_{q} dq_1 \Phi_{q} dq_2.
\end{equation}
$A(M_A,q_1,q_2)$ equals 1 if the system is eclipsing and can mimic a transit and 0 if not, and the mass ratios $q_1\equiv M_B/M_A$ and $q_2 $ (either $M_C/M_A$ or $M_C/M_B$, with 50/50 odds) determine the architecture of the triple system.  $\Phi_q$ is the mass ratio distribution that we used in \S\ref{pappropriate} (flat between 0.1 and 1).  We assign the radius and flux of each component according to the Padova model grids in order to calculate both the non-grazing eclipse probability and the diluted eclipse depths.  Evaluating this integral numerically we obtain $p_a=0.0015$, which results in $\piHT = 0.08 \times 0.0015 = 1.2 \times 10^{-4}$. 

Unlike the blended eclipsing binary scenario, the probability of a target being a hierarchical eclipsing triple does not depend either on galactic latitude or apparent magnitude.  There is a very weak dependence on stellar mass of the primary, but for our calculations we just assume that all target stars have masses close to 1 $M_\odot$, which is reasonable as \Kepler is specifically targeting solar-type stars.





  \begin{figure}[!t]
	\includegraphics[width=3.8in]{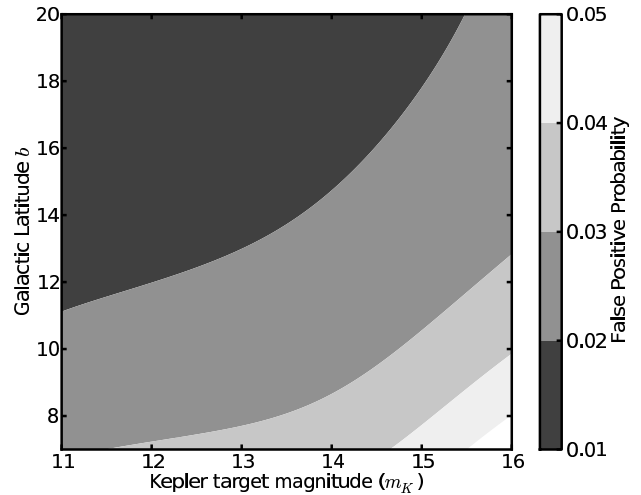}
	\caption{The false positive probability of a \Kepler candidate, according to our basic framework (i.e.~independent of $\delta$), as a function of target star magnitude $m_K$ and galactic latitude.  A planet occurrence rate of 20\% is assumed. This plot assumes that \Kepler is able to internally restrict the radius inside which a possible blended binary might reside to 2$''$.} 
	\label{fig:FPPbasic}
  \end{figure}

\subsection{Basic Framework: Summary and Discussion}
\label{simplesummary}

Now that we have determined the priors for both false positive scenarios, we are able to evaluate the FPP (Equations \ref{eq:simplebayespriorsonly} and \ref{eq:FPP}) by assuming a frequency of close-in planets.  We adopt a 20\% frequency according to the results of the \textit{NASA-UC Eta-Earth Survey} of \citet{howard10}.  This conservative estimate of 20\%, combined with a ~5\% transit probability for a planet on a 10-day orbit (the period we have been assuming up to now) gives $\pi_{\rm pl} = 0.01$.  From a planet detection standpoint, this result is promising, as it gives a 98\% probability that an observed planet-like transit signal around a $m_K=14$ star in the middle of the \Kepler field is authentic, and thus an FPP of only 2\%.  Because of the variation of the background stellar density across the field, this value varies with Galactic latitude and $m_K$, as shown in Figure \ref{fig:FPPbasic}. This is a remarkable result, as it indicates that almost every signal that passes the \Kepler astrometric and photometric false positive tests is likely a planet transit, before any RV confirmation attempts.  

One might rightly pause at this juncture and wonder how the false positive probability for \Kepler can be so low.  After all transit searches up until now, both ground-based (e.g.~HAT, WASP) and space-based (e.g.~CoRoT) been plagued by false positives \citep{OGLEFPs,odonovan06,XOFPs,almenara09}.   To address this, consider what Equation \ref{eq:simplebayespriorsonly} would say about the probability of a transit signal being true for those experiments; these results are summarized in Table \ref{resultstable}.  

Taking the Hungarian-made Automated Telescope Network (HATNet) as an example of a ground-based survey, we note that its 11cm telescopes produce a photometric aperture of about 14$''$ in radius \citep{hartman04}. Using this radius and a depth of 0.5\% as a detection threshold, we repeat the analysis of \S\ref{piBB}, using the line-of-sight simulation at the center of the \Kepler field for the sake of comparison.  For the probability of a possibly blending star to be within the aperture we obtain 1.67, which must obviously now be interpreted as an average number of blending stars per aperture instead of a probability.  For the probability of a blending star to be an appropriate eclipsing binary we obtain $8.4 \times 10^{-4}$, giving $\piBB = 1.67 \times 8.4\times 10^{-4} = 0.0014$.  Following \S\ref{piHT} we calculate $\piHT = 2.7 \times 10^{-4}$.  Finally, taking into account that the probability of a sun-like star hosting a planet easily detectable by this survey is only about 1\%\footnote{for $P < 11.5$d and $M > 0.5 M_J$;  \citet{cumming08}},  then $\pi_{\rm pl} = 0.01 \times 0.05 = 5 \times 10^{-4}$ for this survey.  This results in an FPP of 0.77 for a hot Jupiter-like transit signal for a HAT-like ground-based search, according well with \citet{latham09}, who describe the results of follow-up efforts of a sample of transit candidates, eight of which turned out to be blended binaries and one to be a planet.


The space-based mission CoRoT \citep{baglin03} has also had difficulties with false positives. Though it obtains much better photometric precision than a ground-based search and benefits from uninterrupted observing, its large, 320${\rm ~arcsec}^2$ aperture \citep{almenara09} results in an expected number of 2.71 blended stars for a $m_K=14$ target star.  In addition, its photometric precision is about one part in $10^{3}$, resulting in $\piBB=0.0035$, and $\piHT = 1.4 \times 10^{-4}$.  Assuming then a 20\% occurrence rate of planets detectable by CoRoT, this gives an FPP of 0.27.  At first this appears to somewhat contradict \citet{almenara09}, who reported 6 planets and 25 diluted binaries among CoRoT's ``solved candidates'' (ignoring the ``undiluted binary'' category, as we are not considering that possibility for \Kepler).  However, if one considers how much easier (and faster) it is to identify a false positive than to positively confirm a planet, this prediction can certainly be consistent with these results, as only 49 of their 122 candidates had been solved at the time.  In fact, a prediction of our methods is that many of the unsolved CoRoT candidates are indeed planets.



\begin{figure}[!t]
	\includegraphics[width=3.8in]{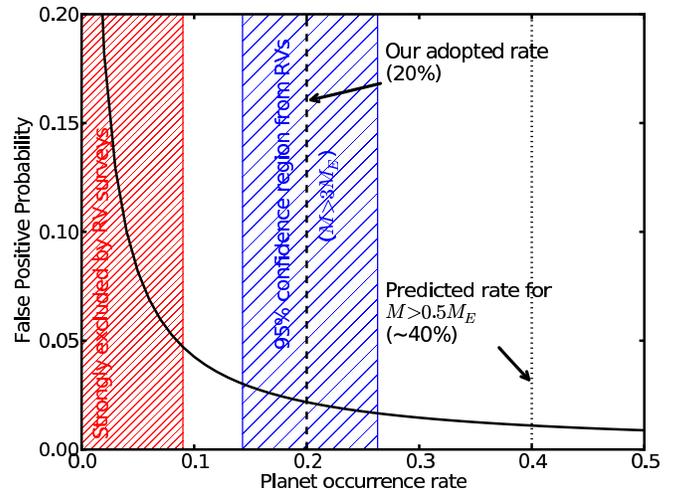}
	\caption{False positive probability as a function of assumed planet occurrence rate, for a $m_K=14$ target star in the center of the \Kepler field.  The occurrence rate of planets detectable by \Kepler is not known for sure, but RV surveys, especially the \textit{NASA-UC Eta-Earth Survey} of \citet{howard10}, have made inroads in measuring the fraction of stars hosting low-mass planets.  The hashed area below 9\% represents the occurrence rate of planets with $P < 50$ days that is ruled out with 95\% confidence by $\eta_{\rm earth}$, counting only the firm detections, and not correcting for completeness.  The central hashed area represents the 95\% confidence region calculated including candidate planets and completeness correction, for minimum masses greater than 3 $M_\oplus$.  Extrapolating their observed mass distribution down to 0.5 $M_\oplus$ brings their total estimated planet occurrence rate to 43\%.  Overall, this plot shows that our derived FPP cannot reasonably be any higher than 5\% if our planet occurrence estimate is incorrect, and will likely be lower.}
	\label{fig:planetfreq}
\end{figure}

Another reasonable question to ask is how uncertainties in our models and assumptions propagate through to uncertainties in FPP.  This is challenging to address exactly, as our analysis rests on the results from TRILEGAL simulations, stellar model grids, and various assumptions about multiple star systems.  Rather than attempt a detailed start-to-finish treatment of all the uncertainties,
we instead investigate what happens if we artificially inject fractional uncertainties into our prior calculations and simulate the results according to our analytic fits.  We find that 20\% fractional uncertainties in background stellar density, appropriate eclipse probability, and hierarchical eclipsing triple probability lead to 17\% fractional uncertainty in FPP.  This is a fiducial example, and the uncertainty in FPP scales linearly with these component uncertainties.

One might also wonder how sensitive our derived FPP for \Kepler is to the assumption that 20\% of stars host planets, as well as how justifiable such an assumption may be.  We address these questions in Figure \ref{fig:planetfreq}.  A 20\% occurrence rate lies in the middle of the measured occurrence rate of planets with minimum masses $> 3 M_\oplus$ and periods $< 50$ days from the \textit{NASA-UC Eta-Earth Survey} of \citet{howard10}.  In addition, even the most pessimistic interpretation of the results from $\eta_{\rm earth}$ allows for a minimum of a 9\% occurrence rate, which would still imply an FPP of only 7\%.  More likely, the true occurrence rate is somewhat higher than our assumption, if not as high as the $\sim$40\% implied by a na\"ive extrapolation of the observed power law-like distribution down to 0.5 $M_\oplus$.  We note that the \textit{NASA-UC Eta-Earth Survey}, as with all RV surveys, is only able to measure minimum masses and thus that the interpretation of the true mass of any individual detection is dependent on an assumption of the overall form of the planet mass function \citep{hoturner}.  However, when an ensemble of minimum mass measurements is available and its distribution resembles a power law with index $\alpha < -1$, the most likely explanation is that the true mass function follows a similar power-law shape.

In summary we may say that several factors contribute to \Kepler being able to minimize the false positive problem compared to previous transit surveys.  First, its ability to astrometrically rule out wide blend scenarios helps mitigate the issue of blended binaries.  Secondly, its photometric precision enables it to identify many false positives based on their secondary eclipses.  And lastly, \Kepler is sensitive to lower-mass planets, which are significantly more common than the larger planets to which ground-based surveys are sensitive.  

\section{Detailed Framework: Considering Transit Depth}
\label{detailed}

 We note that we have not yet discussed any details of the transit signal besides its existence, though some of these details may be important.  For example, one might expect positive blended binaries to be more common at shallower depths (since faint stars are more common than bright stars, and thus more likely to be blended), which might make the BB scenario more of a problem for earth-sized transit signals.  We have also assumed that planets and eclipsing binaries have the same eclipse probability (allowing us to cancel the likelihood factors in Equation \ref{eq:simplebayesLs}), though this is not exactly true either, as both the orbital separations of the systems and the radii of the objects are different.  And finally, for fainter stars and shallower eclipses, it may be more difficult for internal \Kepler procedures to astrometrically identify blends.

With these concerns in mind, we may pursue a more detailed analysis of any particular transit.  There are many features of transit light curves that might all be used in this exercise, but for now we only take into account the depth of the signal, as that is the most easily measured and easily understood quantity.  In this case, Equation \ref{eq:simplebayesLs} becomes:
\begin{equation}
\label{eq:fullbayes}
\Pr({\rm pl} |\delta) = \displaystyle \frac{\mathcal L_{\rm pl}(\delta) \pi_{\rm pl} } { \mathcal L_{\rm pl}(\delta) \pi_{\rm pl} + \mathcal L_{\rm BB}(\delta) \pi_{\rm BB} + \mathcal L_{\rm HT}(\delta) \pi_{\rm HT} }.
\end{equation}
Here the likelihood functions provide a means to quantify the extent to which the conclusions of our simple framework may change as a function of transit depth $\delta$. 


\begin{figure}[!t]
	\includegraphics[width=3.8in]{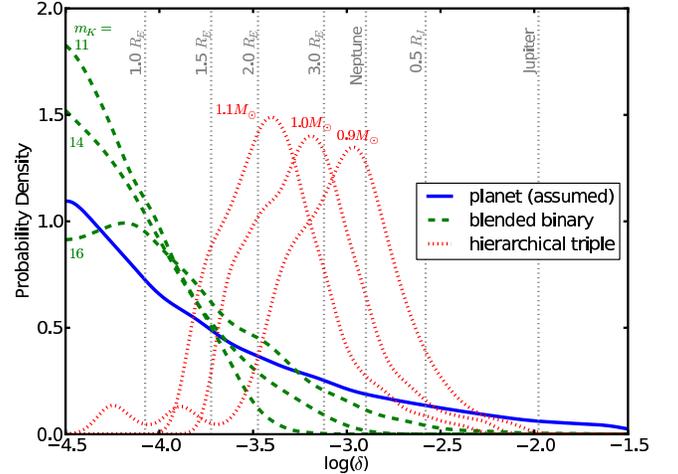}
	\caption{Distributions of apparent ``transit'' depths $\delta$ for different scenarios.  The blended binary and hierarchical triple distributions are based on TRILEGAL simulations with the binary distribution assumptions discussed in \S\ref{framework}.  Examples $\delta$ distributions are given for different target star properties, showing how the blended binary scenario depends on target star apparent magnitude and how the hierarchical triple distribution depends on intrinsic target star mass.  The planet distribution comes from an assumption of a continuous power law in planet radius $dN/dR_p \propto R_p^{-2}$, including random statistical dilution by binary companions.  Note how blended binaries become insignificant for deep transits and how eclipsing triples become insignificant for shallow transits.}
	\label{fig:pdelta}
\end{figure}

\begin{figure}[!t]
	\includegraphics[width=3.8in]{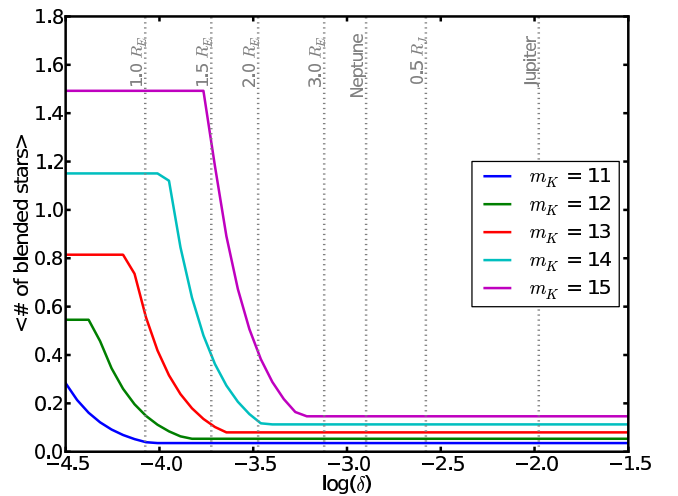}
	\caption{As stars get fainter and transit signals get shallower, the ability for \Kepler to observe a centroid shift indicative of a displaced blended eclipsing binary decreases.  We parametrize this effect according to Equation \ref{eq:rad}.  The plateau towards shallow depths is a result of the maximum blending area for this example being set to an aperture of 8 \Kepler pixels; the location of this plateau for any particular target will depend on its aperture size.  This plot is made according to a galactic latitude in the middle of the \Kepler field; other latitudes will scale appropriately according to the varying stellar density.  The planet radii are marked assuming a Solar-radius star.}
	\label{fig:pblendofd}
\end{figure}

Figure \ref{fig:pdelta} shows the likelihoods that we estimate for the three different scenarios as a function of depth.  The distribution of depths for the blended binary and hierarchical triple scenarios are determined from the same calculations that we used to compute the priors, except rather than just counting all the systems that give depths that are both planetary and detectable, we keep track of the depth of each simulated false positive and build up $\delta$ distributions.  

We calculate the $\delta$ distribution for planets assuming a simple continuous power law distribution of planet radii ($dN/dR_p \propto R_p^{-2}$) between 0.5 and 20 $R_\oplus$, and setting $\delta = (R_p/R_\star)^2$.   While a more sophisticated treatment might involve adopting a planet mass distribution according to RV surveys and theoretical mass-radius relations (e.g.~\citet{fortney07,seager07}), the number of assumptions required for these models and the fact that they do not generally include significant atmospheres for super-Earth-type planets suggests that such efforts are not warranted.  In addition, the current uncertainties in stellar radius of the \Kepler candidate host stars further blur the mapping from $\delta$ to $R_p$.  Thus the main role of the $\delta$ distribution we adopt for planets is to encapsulate the assumption that smaller planets are more common than large ones, which is consistent with radial velocity surveys \citep{howard10}.

Another consideration that should vary with $\delta$ is the ability of \Kepler to astrometrically identify displaced blends.  In \S\ref{framework} we assumed a radius of 2$''$ inside which a blend might reside.  However, this radius should increase as transits get shallower and stars get fainter and the signal-to-noise of the centroid shift signal decreases. This is a question that the \Kepler team should be able to address using simulations of its offset-detecting procedures, but for our purposes we use the radius that the \Kepler team obtained for Kepler 10-b ($1\farcs17$) \citep{batalha11} and assume scaling with $\delta$ and $m_K$ as follows:
\begin{equation}
\label{eq:rad}
r = 1\farcs17 \sqrt{10^{-0.4(11-m_K)}} \left(\frac{\delta}{1.5\times 10^{-4}}\right)^{-1},
\end{equation}
with 11 being the $m_K$ value for Kepler 10. To be conservative we set the minimum $r$ to be $2''$ if this expression gives a smaller value.  On the high end, we cap the radius at $6\farcs4$, corresponding an area equivalent to 8 \Kepler pixels, a typical aperture size (though for any particular target this will vary).  
The square root factor accounts for a diminishing number of photons received as the target star gets fainter, and the inverse relationship with delta is because the centroid shift scales as $\delta$: $\Delta C \sim \delta \cdot r$.  Figure \ref{fig:pblendofd} illustrates this effect; bright stars and deeper transits give $\Pr ({\rm blend})$ as determined in \S\ref{pblend}, but as the target star gets fainter and the signal shallower, the expected number of possibly blending stars begins to increase substantially, up to the point at which our calculated blend radius exceeds the maximum assumed 8-pixel aperture area.  

 \begin{figure}[!t]
	\includegraphics[width=3.8in]{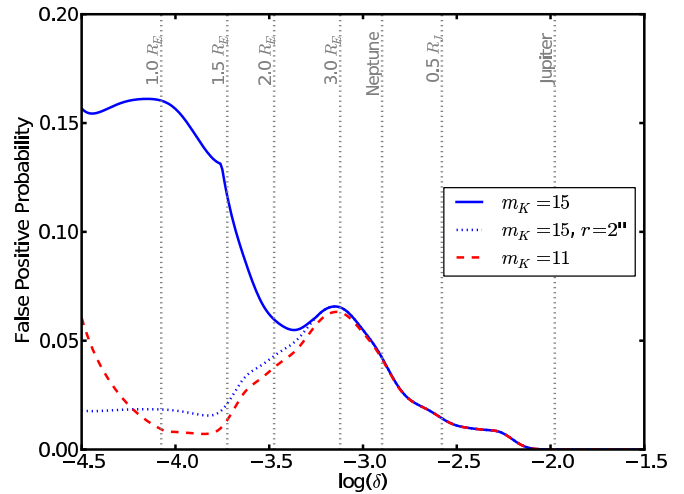}
	\caption{The probability that a signal of a given depth will be a false positive, shown for both an $m_K=11$ and an $m_K=15$ star with Solar properties, Galactic latitude in the center of the \Kepler field, and an 8 pixel \Kepler aperture.  A overall planet occurrence rate of 20\% and a planet radius function $dN/dR \propto R^{-2}$ are assumed.  Note the peak in false positive probability peaks around depths corresponding to about $3 R_\oplus$, due to the peak there in the hierarchical triple $\delta$ distribution.  For the fainter star, the false positive probability increases for shallower transits because it becomes more difficult for \Kepler to rule out displaced blended binaries via astrometry.  However, if a single high-resolution image is able to restrict the possible blend radius to 2$''$, then the FPP for small $\delta$ signals is drastically reduced.  The exact shape of this curve will vary with target star parameters as the shapes of the $\delta$ distributions for the false positive scenarios change (see Figure \ref{fig:pdelta})}
	\label{fig:FPP}
\end{figure}

\begin{figure*}[ht]

\centering
\subfigure[]{
\includegraphics[scale=0.25]{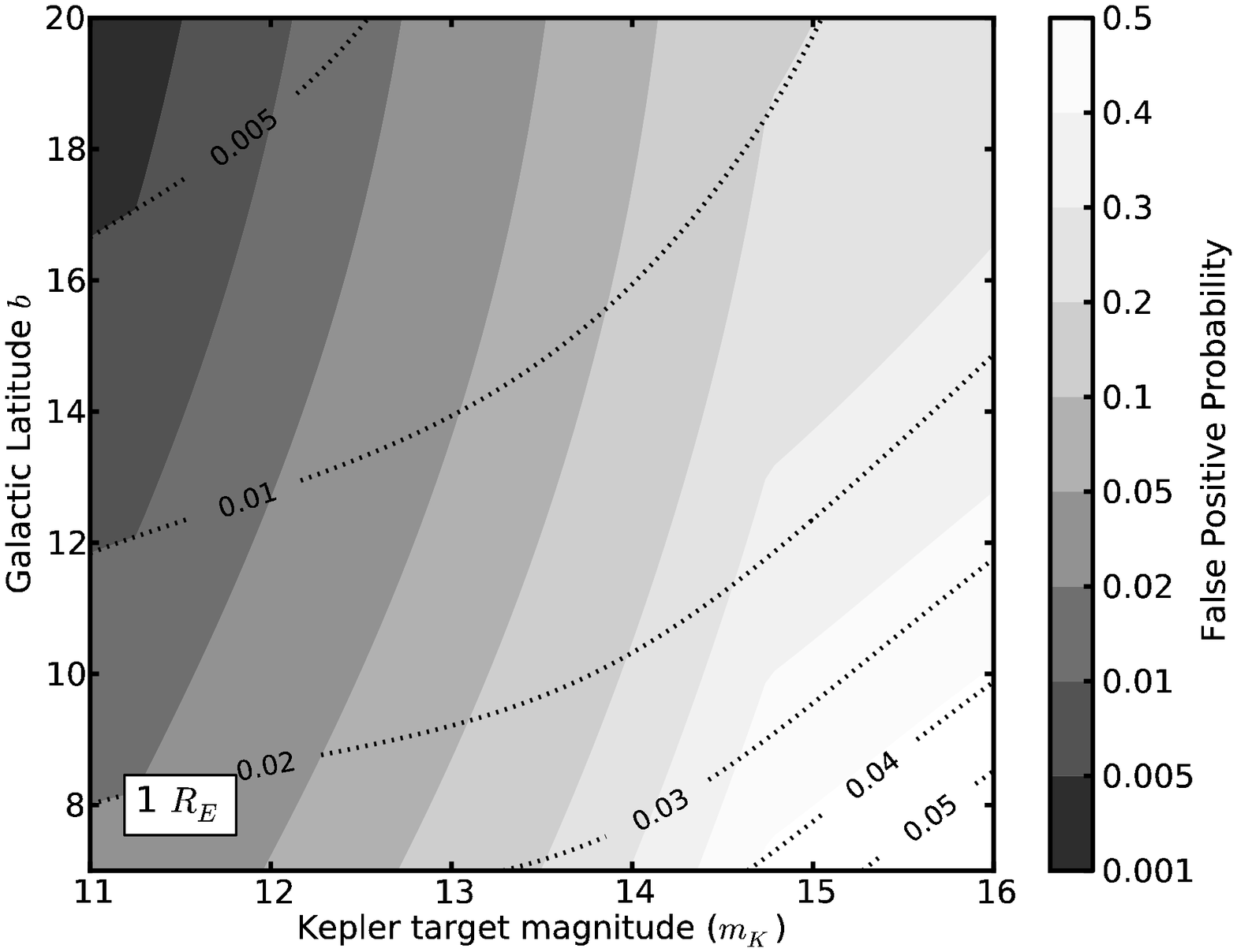}
\label{fig:FPPa}
}
\subfigure[]{
\includegraphics[scale=0.25]{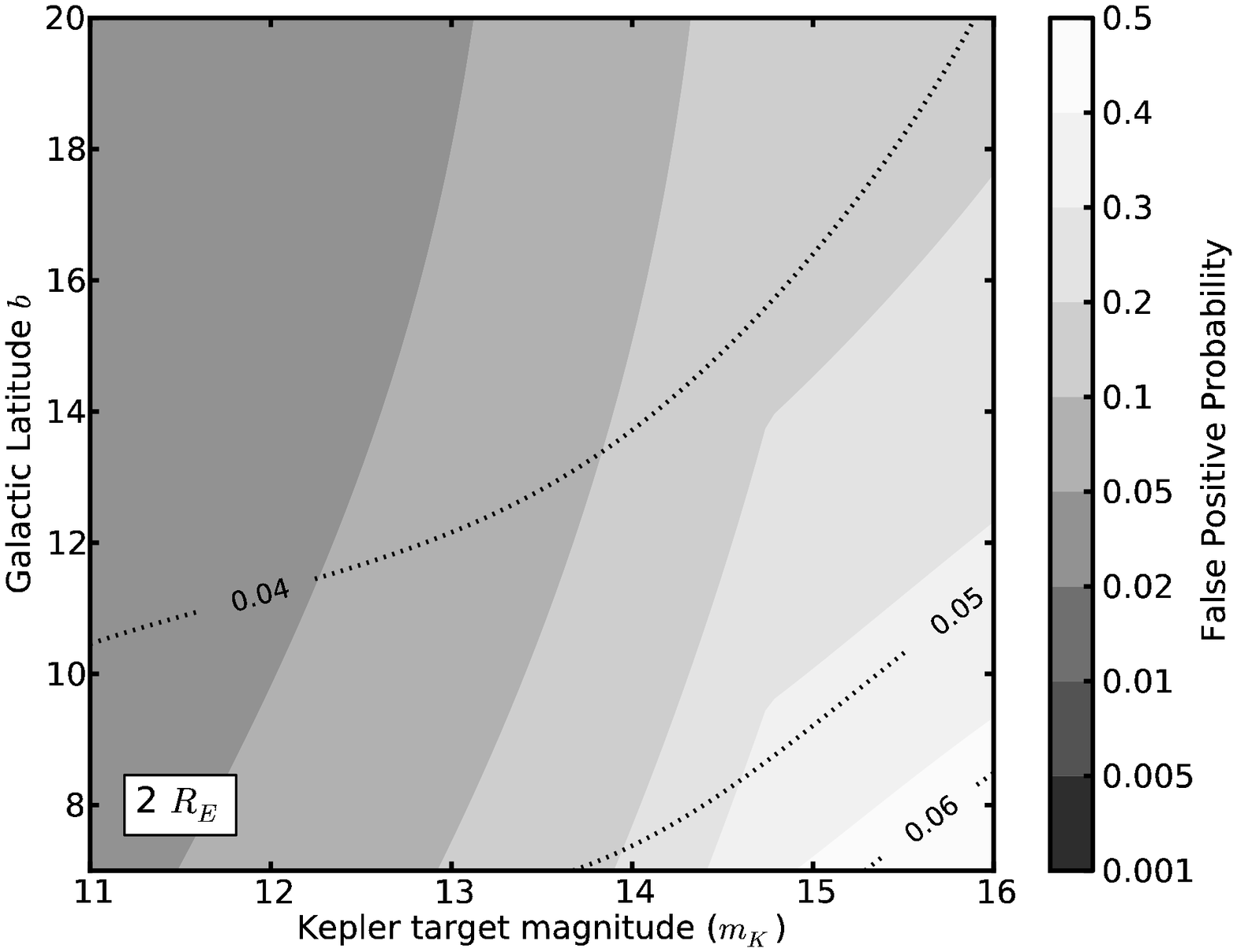}
\label{fig:FPPb}
}
\subfigure[]{
\includegraphics[scale=0.25]{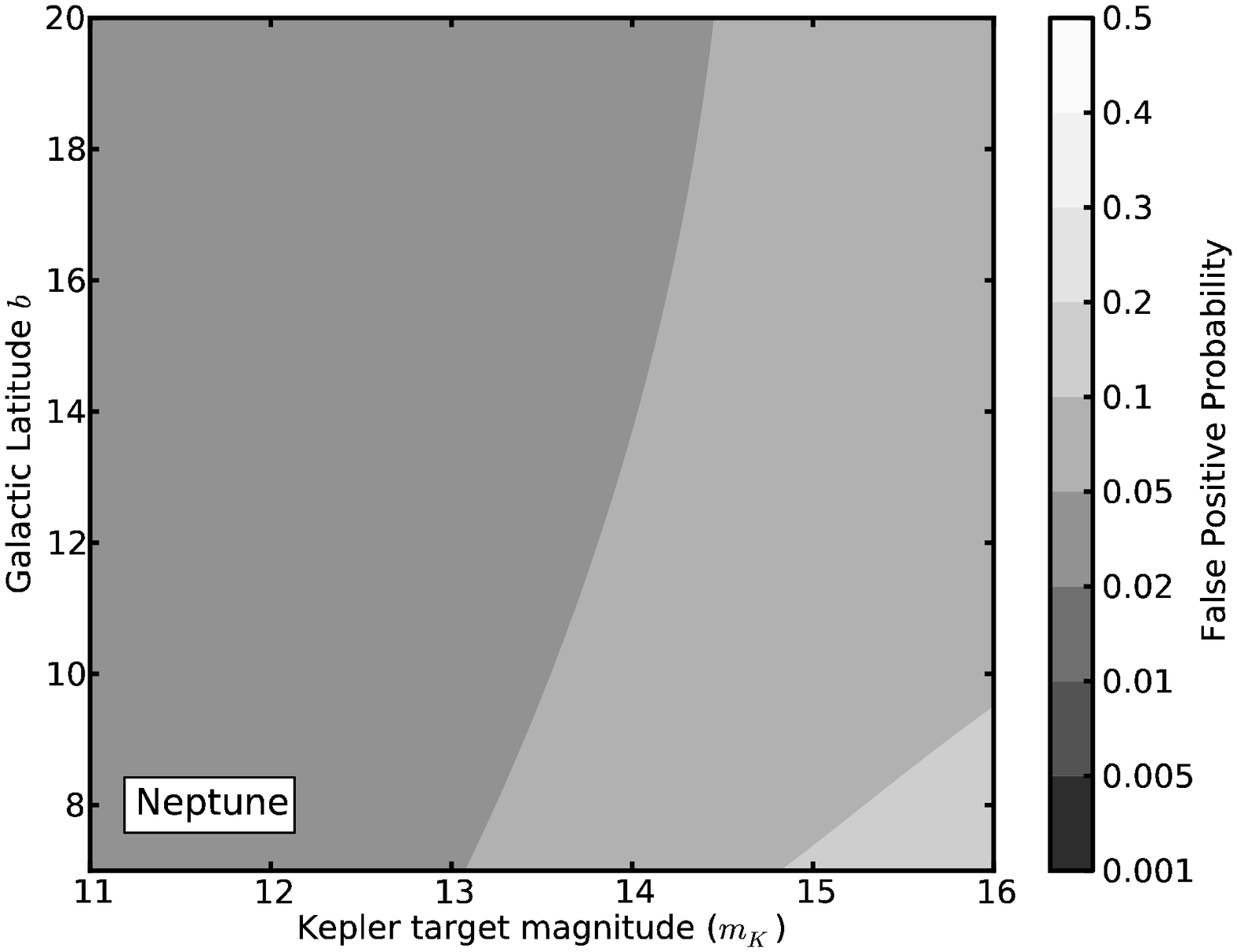}
\label{fig:FPPc}
}

\caption{These plots illustrate the behavior of \Kepler false positive probability (FPP) as a function of target star magnitude ($m_K$) and Galactic latitude, for three particular choices of transit depth $\delta$, all plotted with the same color scale.  A planet occurrence rate of 20\% is assumed, and the target star is fixed to have Solar mass, Solar radius, and a photometric aperture of 8 pixels.  These plots are similar to Figure \ref{fig:FPPbasic} except for they take into account both the changing blend radius as a function of $m_K$ and $\delta$ (Equation \ref{eq:rad}) and the relative likelihoods of false positives and planets at the chosen values of $\delta$.  
All three $\delta$ values show increasing FPP towards fainter target stars and lower galactic latitudes, though the strength of the gradient decreases for the deeper signals, as the relative importance of the hierarchical triple scenario increases.  Dotted lines show the FPP contours if the blend radius were restricted to $2''$, illustrating the power of a single deep high-resolution observation for the shallower signals.  For the Neptune-depth signal, however, as the FPP is dominated by the hierarchical triple scenario, restricting the blend radius to $2''$ has less dramatic an effect (FPP becomes about $4\%$ in this case, and chances very little across the parameter space).
}
\label{fig:FPPsurface}
\end{figure*}

\section{Results}
\label{results}

\subsection{General}
The adoption of these more detailed considerations enables us to estimate the FPP as a function of $\delta$ for a star of given apparent Kepler magnitude, Galactic latitude, stellar radius, stellar mass, and aperture size.  This is illustrated for a fiducial $m_K=14$ Sun-like star in the middle of the \Kepler field (Figure \ref{fig:FPP}), assumed to have an 8-pixel aperture.  We first note that over the whole range of $\delta$, the FPP generally remains low, indicating that these additional considerations do not significantly change the qualitative conclusions we reached within the simple framework.  The majority of transit signals in the \Kepler data release will be actual planets. 

We next draw attention to several features of the plot.  First, we note that any approximately Jupiter-sized candidate, whether around a bright star or faint, is almost certainly a planet.  This is simply because it is extremely difficult to arrange a diluted binary system with a Jupiter-sized primary depth and an undetectable secondary eclipse.  


The second feature of interest is the peak in FPP around $\log \delta = -3.1$, corresponding to about $\sim$3 $R_\oplus$ for a Solar-radius host star.  The origin of this peak may be understood by examining Figure \ref{fig:pdelta} and recognizing that this corresponds to the peak $\delta$ which we predict hierarchical triple false positives to populate for a Solar-mass target star.  This raises the priors-only estimated FPP from \S\ref{simplesummary} by a factor of about 3 for planet candidates slightly smaller than Neptune.  We also note that FPP for signals deeper than this peak is nearly independent of target star apparent magnitude; this is because the eclipsing triple scenario dominates false positives in this regime and the contribution from blended binaries is negligible.


The third significant feature is the rise in FPP towards shallow depths for a target star of Kepler magnitude $m_K=15$, and a similar, though smaller, rise for the brighter $m_K=11$ star at the very shallowest depths.  This is caused by the effect illustrated in Figure \ref{fig:pblendofd}, where the radius outside of which blends may be ruled out by \Kepler astrometry alone should increase with smaller eclipse depth and fainter stars.  Figure \ref{fig:FPP} also illustrates the power of a single deep high-resolution image of any low-amplitude candidate system: any progress in shrinking the radius inside which a blended binary might reside will significantly decrease the FPP for Earth-sized transit signals, under the assumption that the occurrence rate of planets rises toward smaller masses, as we have assumed.  

We note that the plots in Figures \ref{fig:pdelta}, \ref{fig:pblendofd} and \ref{fig:FPP} are only for particular chosen values of magnitude and a single Galactic latitude in the middle of the \Kepler field, as well as for particular choices of stellar properties.  We present a more comprehensive illustration of the FPP manifold in Figure 8, 
choosing three specific values of $\delta$ to illustrate how the FPPs for different types of signals vary with target star magnitude and Galactic latitude.   We fix the target star to have Solar properties in these examples. 

Earth-sized transits show a steep gradient across the field and towards fainter stars; this is a result of increasing contribution to the FPP from blended binaries (see Figure \ref{fig:pblendbasic}), combined with the increased blend radius for a shallow transit (Figure \ref{fig:pblendofd}).  This gradient is shallower for a $2 R_\oplus$ signal and almost disappears for a Neptune-sized signal, because of the growing contribution of the hierarchical triple scenario.  These plots also illustrate the potential power of deep high-resolution imaging follow-up observations.  If such an image is taken and no companion is found outside a radius of a few arcseconds, then that dramatically reduces the FPP for shallow signals, as illustrated with the dotted contours in Figure 8.


\begin{deluxetable*}{ccccccccccc|c}
	\tabletypesize{\footnotesize}
	\tablecolumns{12}
	\tablecaption{False Positive Probabilities for Kepler Planet Candidates\tablenotemark{1}}
	\tablehead{\colhead{KOI} &  \colhead{$\delta$} & \colhead{$m_K$} & \colhead{\# pixels} & \colhead{$b$}  & \colhead{$P$} &  \colhead{$M_\star$} & \colhead{$R_\star$} & \colhead{$L_{\rm BB}$} & \colhead{$L_{\rm HT}$} & \colhead{$L_{\rm pl}$} & \colhead{FPP} \\ (1) & (2) & (3) & (4) & (5) & (6) & (7) & (8) & (9) & (10) & (11) & (12) }
	\startdata
	371.01 & 1.11e-03 & 12.19 & 19 & 5.94 & 278.000 & 1.33 & 3.01 & 2.4e-07 & 1.4e-06 & 0.00023 & 0.01 \\
	372.01 & 7.64e-03 & 12.39 & 12 & 6.82 & 125.612 & 1.05 & 0.95 & 1.5e-23 & 4.5e-09 & 0.00014 & $<0.01$ \\
	373.01 & 5.97e-04 & 12.77 & 8 & 11.79 & 135.194 & 1.11 & 1.30 & 1.2e-06 & 2.1e-05 & 0.0005 & 0.04 \\
	374.01 & 5.95e-04 & 12.21 & 13 & 13.68 & 172.673 & 1.11 & 1.26 & 2.8e-07 & 1.9e-05 & 0.00042 & 0.04 \\
	375.01 & 4.70e-03 & 13.29 & 13 & 15.91 & 220.000 & 1.07 & 1.04 & 3.9e-10 & 8.3e-07 & 0.00012 & 0.01 \\
	377.01 & 6.94e-03 & 13.80 & 6 & 14.49 & 19.258 & 1.00 & 0.68 & 1.5e-11 & 1.5e-06 & 0.00052 & $<0.01$ \\
	377.02 & 6.24e-03 & 13.80 & 6 & 14.49 & 38.912 & 1.00 & 0.68 & 1.1e-08 & 1.5e-06 & 0.00034 & $<0.01$ \\
	377.03 & 2.25e-04 & 13.80 & 6 & 14.49 & 1.593 & 1.00 & 0.68 & 0.00023 & 0.00025 & 0.016 & 0.03 \\
	379.01 & 2.51e-04 & 13.32 & 10 & 9.61 & 6.717 & 1.19 & 1.59 & 8.6e-05 & 0.00026 & 0.0059 & 0.06 \\
	384.01 & 1.76e-04 & 13.28 & 8 & 8.46 & 5.080 & 1.09 & 1.22 & 0.00031 & 0.00015 & 0.0083 & 0.05 \\
	385.01 & 2.69e-04 & 13.44 & 5 & 9.85 & 13.146 & 1.04 & 1.04 & 4.5e-05 & 9.3e-05 & 0.0035 & 0.04 \\
	386.01 & 8.45e-04 & 13.84 & 5 & 8.61 & 31.158 & 1.11 & 1.12 & 1.9e-06 & 3.2e-05 & 0.0011 & 0.03 \\
	386.02 & 6.60e-04 & 13.84 & 5 & 8.61 & 76.735 & 1.11 & 1.12 & 2.9e-06 & 2.8e-05 & 0.00072 & 0.04 \\
	387.01 & 9.41e-04 & 13.58 & 9 & 13.50 & 13.900 & 0.69 & 0.74 & 2.9e-06 & 1.1e-05 & 0.0018 & 0.01 
	\enddata
	\tablenotetext{1}{Here is printed only a portion of the table to show its format and contents; all 1235 candidates are listed in the full version of the table, available online at exoplanets.org/data/KOIFPPtable.txt.}
	\tablenotetext{(1)}{KOI identifier, from \citet{borucki11}}
	\tablenotetext{(2)}{transit depth}
	\tablenotetext{(3)}{Kepler magnitude}
	\tablenotetext{(4)}{Size, in \Kepler pixels ($4''$ square each) of the photometric aperture, according to the publicly available pixel data.}
	\tablenotetext{(5)}{Galactic latitude of target star, in degrees}
	\tablenotetext{(6)}{Period of candidate, in days}
	\tablenotetext{(7)}{Stellar mass, according to the Kepler Input Catalog (KIC)}
	\tablenotetext{(8)}{Stellar radius, according to the KIC}
	\tablenotetext{(9)}{Likelihood $\times$ prior for the blended binary scenario}
	\tablenotetext{(10)}{Likelihood $\times$ prior for the eclipsing hierarchical triple scenario}
	\tablenotetext{(11)}{Likelihood $\times$ prior for the transiting planet}
	\tablenotetext{(11)}{False positive probability $= 1 - L_{\rm pl}/(L_{\rm pl} + L_{\rm BB} + L{\rm HT})$}
	\label{table:KOIresults}
\end{deluxetable*}

\begin{figure}[!t]
	\includegraphics[width=3.8in]{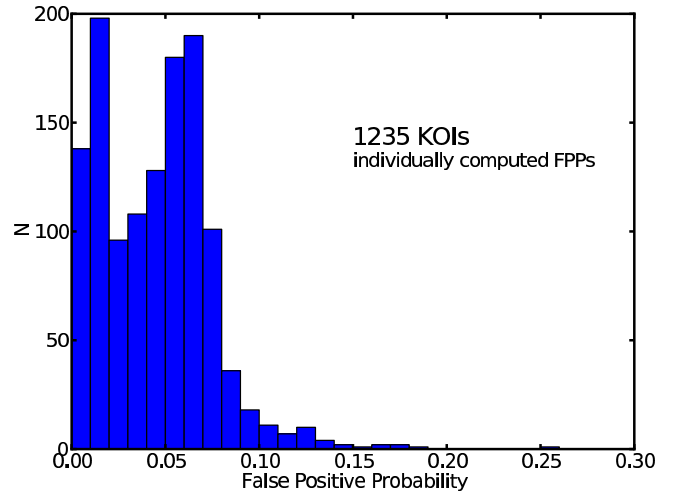}
	\caption{The distribution of false positive probabilities (FPPs) among the 1235 \Kepler planet candidates announced in \citet{borucki11}.  FPP for each candidate is calculated individually, taking into account the apparent Kepler magnitude, Galactic latitude, mass and radius of the host star, the depth of the signal, and the number of pixels contained the optimal aperture used for \Kepler photometry.  Almost all (1193) have FPPs less than 10\%, and over half (668) have FPPs less than 5\%.  The mean FPP of the sample is 4\%, indicating that we expect there to be fewer than 50 false positives among the candidate sample.}
	\label{fig:FPPhist}
\end{figure}

\subsection{Application to Kepler Candidates}
We apply the framework discussed above to calculate the FPP for every \Kepler Object of Interest (KOI) published in \citet{borucki11}; these results are summarized in Table \ref{table:KOIresults}.  For each KOI we generate individualized $\delta$ distributions for the different false positive scenarios using the relevant Kepler magnitude, Galactic latitude, stellar parameters from the Kepler Input Catalog.  We then calculate the FPP using the actual area of the photometric aperture, as determined from the publicly available pixel data for each KOI and the transit depth as given in \citet{borucki11}.  The distribution of FPPs is illustrated in Figure \ref{fig:FPPhist}.

In Table \ref{table:KOIresults} we list the KOI parameters relevant to the FPP calculation, the calculated FPPs, and the values of the intermediate factors in the calculation, which we summarize as $L_{\rm pl}$, $L_{\rm BB}$ and $L_{\rm HT}$, where
\begin{equation}
\label{eq:intermediateresults}
L_{\rm pl} = \mathcal L_{\rm pl}(\delta) \pi_{\rm pl} = f_{\rm pl} \cdot {\rm Pr(Transit)} \cdot \Phi_{\rm pl}(\log \delta),
\end{equation}
where $f_{\rm pl}$ is the overall planet occurrence frequency, Pr(Transit) is the geometric transit probability, and $\Phi_{\rm pl} = dN/d\log \delta$ is the probability density function for $\log \delta$.  Thus 
\begin{equation}
\label{eq:FPPagain}
\rm{FPP} = 1-\frac{L_{\rm pl}}{L_{\rm pl} + L_{\rm BB} + L_{\rm HT}},
\end{equation}
where $L_{\rm BB}$ and $L_{\rm HT}$ are the corresponding terms for the two false positive scenarios.

We list these individual components in the table primarily because the FPP calculation fundamentally depends on assumptions of the planetary occurrence rate and radius distribution, and different assumptions will result in different FPPs.   Though we show in Figure \ref{fig:planetfreq} that these assumptions are unlikely to dramatically affect the final FPP numbers, one could in principle calculate $L_{\rm pl}$ based on different assumptions and recalculate FPP, given all the components.


\section{Discussion: Relationship to ``BLENDER''} 
\label{discussion}

The FPP analysis we present in this paper is not the first false positive analysis that has been done regarding \Kepler candidates.  In fact, the \Kepler team has statistically ``validated'' three planets: Kepler-9d \citep{torres11}, Kepler-11g \citep{lissauer11}, and Kepler-10c \citep{fressin11} by demonstrating that the chance of any of those signals being due to a false positive is low enough to consider the candidate a \textit{bona fide} planet.  This has been done using the procedure the \Kepler team has named BLENDER.

BLENDER attempts to directly model the candidate light curve using every conceivable false positive scenario, informed by high-resolution imaging follow-up observations.  The goodness-of-fit of the false positive models is then compared to the best-fit planetary model.  The false positive scenarios that cannot fit the light curve as well as as a transiting planet model are rejected.   The \textit{a priori} likelihood of the remaining scenarios (those false positive scenarios that provide comparable-quality fits to the light curve) is then assessed relative to the likelihood of a \textit{bona fide} transiting planet, and if the planetary explanation is much more likely, then the planet is considered validated.

As can be inferred from the fact that the \Kepler team has published only three validated planets to date out of over 1200 planet candidates that have been made public, BLENDER is a very time-consuming procedure, being both computationally expensive and labor-intensive.   Relying on extensive modeling of individual light curves and requiring a suite of follow-up observations to be most effective, it can only be applied to single KOIs on an individual basis.  

If BLENDER may be characterized as a ``deep and narrow'' false positive analysis tool, the FPP analysis we present in this paper might be described as ``shallow and wide.''  It takes only 15 seconds per candidate for us to generate the $\delta$-distributions required to calculate the individualized FPP numbers listed in Table \ref{table:KOIresults}, which makes our analysis easily and immediately applicable to all the KOIs, whereas BLENDER takes months of computation and analysis per candidate.  On the other hand, BLENDER takes into account all possible information about each KOI (detailed light curve shape, AO imaging, multiwavelength transit information, etc.), whereas we only consider the depth of the transit signal and the properties of the target star.

Another way to think of the relationship between our FPP analysis and BLENDER is that if BLENDER is a $N$-step procedure, our analysis is step $N$.  We ignore most of the detail of the light curve and make no use of any follow-up observations, but go straight to the \textit{a priori} likelihood calculation and do that step as carefully as possible.  What is remarkably encouraging for the \Kepler mission is that even this ``shallow,'' single-step analysis is enough to determine that the false positive probability for almost every KOI is less than 10\%, and for over half the KOIs is less than 5\%.   

If our analysis is step $N$ of the BLENDER process, how would the first $N-1$ steps be incorporated into the present analysis to improve upon the FPPs published here? First, consider that if $x = L_{\rm pl}$ and $y = L_{\rm BB} + L_{\rm HT}$, then the probability $p_{\rm pl}$ that a signal is a planet is the following:
\begin{equation}
p_{\rm pl} = \frac{x}{x+y}.
\end{equation}
This may be rewritten as
\begin{equation}
p_{\rm pl} = \frac{1}{1+y/x}.
\end{equation}
If $y \ll x$ (as we have shown it typically is) then 
\begin{equation}
p_{\rm pl} \approx 1 - \frac{y}{x},
\end{equation}
or 
\begin{equation}
{\rm FPP} \approx \frac{y}{x}.
\end{equation}
The typical role of BLENDER in this context can then be thought of as multiplying $y$ by a factor we call $f_{\rm BLENDER}$ ($0 < f_{\rm BLENDER} < 1$) that represents the fraction of the potential false positive scenarios (weighted by their intrinsic likelihoods) that produce acceptable fits to the light curve.  Thus if BLENDER were to rule out 90\% of the false positive scenarios considered in our analysis ($f_{\rm BLENDER} = 0.1$) for a particular system, then that would decrease FPP for that system by a factor of 10---such analysis would be enough to make ${\rm FPP} < 0.01$ for almost every KOI.
  
In a similar spirit, for those KOIs whose FPP is dominated by the blended binary scenarios (mostly the shallowest signals), $y$ can also be significantly decreased simply if deep high-resolution imaging shows no potentially blending companions.  This effect is illustrated in Figure 8, where dotted FPP contours are drawn illustrating the effect of restricting the ``blend radius'' to 2$''$.  Decreasing this even further to 1$''$ or smaller would give another factor of 4 or more reduction in FPP.   Thus we demonstrate that simply obtaining deep high-resolution images may be just as effective as the entire BLENDER analysis for probabilistically validating some KOIs!
  
In some cases of course, follow-up imaging observations will identify the presence of nearby stars within the ``blend radius'' inside of which astrometric offset methods were previously unable to identify displaced blends.  In these cases, the analysis presented in this paper must be superceded by a more specifically tailored analysis such as BLENDER.  In general a detected nearby blend will cause the prelimininary FPP to substantially increase, as the $\rm{Pr(Blend)}$ factor that we found to be of order $\sim$$0.10$ (\S\ref{pblend}) is then divided out from the $L_{\rm BB}$ term, making it more comparable to the $L_{\rm pl}$ term.  In these cases a full suite of follow-up observations and the more detailed wholistic approach that BLENDER utilizes will become necessary to validate candidates.


\section{Caveats and Conclusions}
\label{conclusions}
We present both a framework to analyze the \textit{a priori} false positive probability (FPP) of \Kepler planet candidates and preliminary FPPs for the entire sample of 1235 released candidates, finding that ${\rm FPP} <10\%$ for almost all the KOIs and $< $$5\%$ for over half the KOIs.  The philosophy we adopt in this work is to calculate conservative upper limits to these FPPs; further analysis may well demonstrate them to be lower, but we do not expect them to be higher.  Thus we may say confidently say that our analysis indicates that fewer than 50 of the 1235 candidates are likely to turn out to be false positives. 

However, these conclusions are based on several assumptions that come with some caveats:
\begin{itemize}
\item We assume that all candidates have passed all preliminary false-positive-vetting procedures that are possible using \Kepler photometry and astrometry alone.  In particular we assume that the transits are not obviously V-shaped, there is no detectable secondary eclipse, and that careful centroid analysis has not revealed the presence of a displaced blended binary.  If photometry or astrometry for a candidate actually does turn out to indicate a possible false positive, then the FPPs calculated in this paper for that KOI are not accurate.
\item We assume host star stellar parameters according to the Kepler Input Catalog (KIC).  If stellar radii or stellar types are found to be significantly different from the KIC estimates, then that could change the interpretation of transit signals (e.g.~turning a Jupiter-sized planet into an M-dwarf).  
\item We assume a planet radius function that increases towards smaller planets.  There are many reasons, both theoretical and observational, to assume this is correct, but if it is not, then the false positive numbers for the smallest candidates would be a factor of two or so higher.
\end{itemize}

We also emphasize that the intention of this paper is not to encourage other analyses to completely ignore the possibility that some \Kepler candidates might be false positives.  Rather, we suggest that in statistical analyses using the ensemble of KOIs to investigate the distribution of planet properties, the FPPs in this paper (or based on the calculations in this paper; e.g.~with different assumptions of the planet occurrence rate or radius function) be used to count ``fractional planets''; i.e.~for a KOI with ${\rm FPP} = 0.05$ to count as 95\% of a planet.

Finally, we provide several suggestions to guide and optimize \Kepler follow-up efforts, based on the results of our analysis:
\begin{itemize}
\item For the shallowest candidates, or those for which a blended binary is the most likely false positive scenario, we recommend deep high-resolution imaging (with a target contrast ratio corresponding to the depth of the signal:  $\Delta m_K = -2.5 \log \delta$), as excluding the presence of potentially blending stars at close separation will be the quickest path toward validation of such systems.  Contrast ratios up to 10 magnitudes as close as 1$''$ have long proven to be technically feasible \citep[e.g.]{luhman02,biller07}.
\item For candidates of intermediate depth for which a hierarchical eclipsing triple is the most likely false positive scenario we recommend follow-up efforts targeted toward the identification of physically bound companions to the KOI.   High-resolution imaging is one useful tool here (though not necessarily as deep as those observations targeting projected binaries) to target wide-separation companions, but high S/N spectroscopy (both optical and infrared) may be even more important, in order to spectroscopically identify or constrain the presence of low-mass stellar companions.
\item For the candidates with the largest implied radii we recommend primarily spectroscopic follow-up to improve our knowledge of the physical parameters of the candidate host stars, in order to rule out the possibility of an eclipsing binary being misclassified as a transiting planet due to an incorrect assumed radius.
\end{itemize}

In summary, the exquisite photometric and astrometric precision of the
\Kepler instrument enables many of the false positives that have traditionally
plagued transit surveys to be identified prior to follow-up observations. The
result is that the majority of the candidates announced by \citet{borucki11}
are likely to be \textit{bona fide} planets. Thus, having surveyed the landscape of
false positives in the \Kepler field, we conclude that the outlook is bright for
statistical analyses of exoplanet occurrence and properties based on the
data made public by the \Kepler team.

\acknowledgments{We gratefully acknowledge Ed Turner and Scott Gaudi for their
thoughtful and detailed comments on earlier drafts of this paper. 
We also thank Geoff Marcy, Steve Bryson, and Guillermo Torres 
for their useful discussions related to the \Kepler mission at the 
January 2011 AAS meeting and during the referee process. We thank Jessica Lu for bringing 
TRILEGAL to our attention. Finally, we acknowledge the 
dedication and hard work of the \Kepler team for opening up this
amazing new frontier in exoplanetary science. }

\appendix
\section{Blended Planets}

In the present work we consider as false positives only astrophysical configurations that do not involve any planets but still mimic the signal of a transiting planet.  However, there are various other scenarios involving ``blended planets'' that, while not strictly false positives (i.e. a transiting planet is still involved), may contribute significantly to uncertainty in the planet parameters derived from the transit signal.  A ``blended planet'' for our purposes is a transit signal that appears to be a planet of a particular size transiting the target star but is actually a larger planet transiting a fainter blended star.  As before, these scenarios can be divided into chance-alignment systems or physically associated hierarchical systems.

We have calculated that chance-alignment blended planets are significantly less likely to occur than their blended stellar binary cousins; this can be heuristically understood from the following considerations:

\begin{itemize}
\item Because the deepest intrinsic planetary transits have depths of only $\sim$$0.02$ and the diluted signal has to be detectable (we have adopted $\delta \gtrsim 10^{-4}$ as a threshold), then 
the maximum contrast between the target star and the blending star is $\Delta m_K = 5.75$, which is significantly less than the $\Delta m_K = 10$ we adopted for blended binaries in \S\ref{pblend}.  The sky density of stars available for the chance-alignment blended planet scenario is thus about 5.5 times lower than that for the blended binary scenario, according to the TRILEGAL simulations.

\item Our assumed planet frequency ($\sim$$20\%$) is lower than our assumed binary fraction ($\sim$$40\%$).

\item The largest planets, while the most amenable to causing the blended planet scenario because of their larger intrinsic transit depth, are the least common---only $\sim$$1\%$ of solar-type stars host close-in giant planets, and this occurrence rate is even lower for lower-mass stars \citep{endl03,johnson10}, which are the most common blending stars.

\end{itemize}

Physically associated hierarchical planets, on the other hand, might well be relatively common compared to the stellar false positive scenarios or chance-alignment blended planets.  
Another way of saying this is that binary stellar systems are relatively common, and so it seems likely that a substantial fraction of \Kepler targets (and therefore candidates) are in fact binaries of unknown architecture.  The net effect of this on the interpretation of the sample of planet candidates will be additional uncertainty in the derived planet properties due to both diluting light from a binary companion and from possible stellar misclassification by the \Kepler Input Catalog, which assumes each star is single.  We note that the \Kepler team does include blended planets in the BLENDER procedure, and in fact that such scenarios are often the most difficult to rule out (\Kepler team, 2011, private comm.).

In summary, while the analysis presented in this paper may provide confidence that ``classic false positive'' stellar systems are not often masquerading as \Kepler transiting planet candidates, we do caution that uncertainties regarding candidate host systems (including whether or not they are binary) must be considered in any statistical analysis of the whole candidate sample.  

\bibliographystyle{apj}
\bibliography{myrefs}

\begin{thebibliography}{42}
\expandafter\ifx\csname natexlab\endcsname\relax\def\natexlab#1{#1}\fi

\bibitem[{REV(????)}]{REVTEX41Control}
 ????

\bibitem[{08(1)}]{apsrev41Control}
08. 1

\bibitem[{{Almenara} {et~al.}(2009){Almenara}, {Deeg}, {Aigrain}, {Alonso},
  {Auvergne}, {Baglin}, {Barbieri}, {Barge}, {Bord{\'e}}, {Bouchy}, {Bruntt},
  {Cabrera}, {Carone}, {Carpano}, {Catala}, {Csizmadia}, {de La Reza},
  {Deleuil}, {Dvorak}, {Erikson}, {Fridlund}, {Gandolfi}, {Gillon}, {Gondoin},
  {Guenther}, {Guillot}, {Hatzes}, {H{\'e}brard}, {Jorda}, {Lammer},
  {L{\'e}ger}, {Llebaria}, {Loeillet}, {Magain}, {Mayor}, {Mazeh}, {Moutou},
  {Ollivier}, {P{\"a}tzold}, {Pont}, {Queloz}, {Rauer}, {R{\'e}gulo}, {Renner},
  {Rouan}, {Samuel}, {Schneider}, {Shporer}, {Wuchterl}, \&
  {Zucker}}]{almenara09}
{Almenara}, J.~M., {et~al.} 2009, \aap, 506, 337

\bibitem[{{Baglin}(2003)}]{baglin03}
{Baglin}, A. 2003, Advances in Space Research, 31, 345

\bibitem[{{Basri} {et~al.}(2010){Basri}, {Walkowicz}, {Batalha}, {Gilliland},
  {Jenkins}, {Borucki}, {Koch}, {Caldwell}, {Dupree}, {Latham}, {Meibom},
  {Howell}, \& {Brown}}]{basri10}
{Basri}, G., {et~al.} 2010, \apjl, 713, L155

\bibitem[{{Batalha} {et~al.}(2010{\natexlab{a}}){Batalha}, {Rowe}, {Gilliland},
  {Jenkins}, {Caldwell}, {Borucki}, {Koch}, {Lissauer}, {Dunham}, {Gautier},
  {Howell}, {Latham}, {Marcy}, \& {Prsa}}]{batalha10b}
{Batalha}, N.~M., {et~al.} 2010{\natexlab{a}}, \apjl, 713, L103

\bibitem[{{Batalha} {et~al.}(2010{\natexlab{b}}){Batalha}, {Borucki}, {Koch},
  {Bryson}, {Haas}, {Brown}, {Caldwell}, {Hall}, {Gilliland}, {Latham},
  {Meibom}, \& {Monet}}]{batalha10a}
---. 2010{\natexlab{b}}, \apjl, 713, L109

\bibitem[{{Batalha} {et~al.}(2011){Batalha}, {Borucki}, {Bryson}, {Buchhave},
  {Caldwell}, {Christensen-Dalsgaard}, {Ciardi}, {Dunham}, {Fressin},
  {Gautier}, {Gilliland}, {Haas}, {Howell}, {Jenkins}, {Kjeldsen}, {Koch},
  {Latham}, {Lissauer}, {Marcy}, {Rowe}, {Sasselov}, {Seager}, {Steffen},
  {Torres}, {Basri}, {Brown}, {Charbonneau}, {Christiansen}, {Clarke},
  {Cochran}, {Dupree}, {Fabrycky}, {Fischer}, {Ford}, {Fortney}, {Girouard},
  {Holman}, {Johnson}, {Isaacson}, {Klaus}, {Machalek}, {Moorehead},
  {Morehead}, {Ragozzine}, {Tenenbaum}, {Twicken}, {Quinn}, {VanCleve},
  {Walkowicz}, {Welsh}, {Devore}, \& {Gould}}]{batalha11}
---. 2011, \apj, 729, 27

\bibitem[{{Beatty} \& {Gaudi}(2008)}]{beatty08}
{Beatty}, T.~G., \& {Gaudi}, B.~S. 2008, \apj, 686, 1302

\bibitem[{{Biller}(2007)}]{biller07}
{Biller}, B.~A. 2007, PhD thesis, The University of Arizona

\bibitem[{{Borucki} {et~al.}(2008){Borucki}, {Koch}, {Basri}, {Batalha},
  {Brown}, {Caldwell}, {Christensen-Dalsgaard}, {Cochran}, {Dunham}, {Gautier},
  {Geary}, {Gilliland}, {Jenkins}, {Kondo}, {Latham}, {Lissauer}, \&
  {Monet}}]{borucki08}
{Borucki}, W., {et~al.} 2008, in IAU Symposium, Vol. 249, IAU Symposium, ed.
  {Y.-S.~Sun, S.~Ferraz-Mello, \& J.-L.~Zhou}, 17--24

\bibitem[{{Borucki} {et~al.}(2011){Borucki}, {Koch}, {Basri}, {Batalha},
  {Brown}, {Bryson}, {Caldwell}, {Christensen-Dalsgaard}, {Cochran}, {DeVore},
  {Dunham}, {Gautier}, {Geary}, {Gilliland}, {Gould}, {Howell}, {Jenkins},
  {Latham}, {Lissauer}, {Marcy}, {Rowe}, {Sasselov}, {Boss}, {Charbonneau},
  {Ciardi}, {Doyle}, {Dupree}, {Ford}, {Fortney}, {Holman}, {Seager},
  {Steffen}, {Tarter}, {Welsh}, {Allen}, {Buchhave}, {Christiansen}, {Clarke},
  {D{\'e}sert}, {Endl}, {Fabrycky}, {Fressin}, {Haas}, {Horch}, {Howard},
  {Isaacson}, {Kjeldsen}, {Kolodziejczak}, {Kulesa}, {Li}, {Machalek},
  {McCarthy}, {MacQueen}, {Meibom}, {Miquel}, {Prsa}, {Quinn}, {Quintana},
  {Ragozzine}, {Sherry}, {Shporer}, {Tenenbaum}, {Torres}, {Twicken}, {Van
  Cleve}, \& {Walkowicz}}]{borucki11}
{Borucki}, W.~J., {et~al.} 2011, ArXiv e-prints

\bibitem[{{Brown}(2003)}]{brown03}
{Brown}, T.~M. 2003, \apjl, 593, L125

\bibitem[{{Chabrier}(2001)}]{chabrier01}
{Chabrier}, G. 2001, \apj, 554, 1274

\bibitem[{{Chabrier} {et~al.}(2000){Chabrier}, {Baraffe}, {Allard}, \&
  {Hauschildt}}]{chabrier00}
{Chabrier}, G., {Baraffe}, I., {Allard}, F., \& {Hauschildt}, P. 2000, \apj,
  542, 464

\bibitem[{{Cumming} {et~al.}(2008){Cumming}, {Butler}, {Marcy}, {Vogt},
  {Wright}, \& {Fischer}}]{cumming08}
{Cumming}, A., {Butler}, R.~P., {Marcy}, G.~W., {Vogt}, S.~S., {Wright}, J.~T.,
  \& {Fischer}, D.~A. 2008, \pasp, 120, 531

\bibitem[{{Endl} {et~al.}(2003){Endl}, {Cochran}, {Tull}, \&
  {MacQueen}}]{endl03}
{Endl}, M., {Cochran}, W.~D., {Tull}, R.~G., \& {MacQueen}, P.~J. 2003, \aj,
  126, 3099

\bibitem[{{Evans} \& {Sackett}(2010)}]{evans10}
{Evans}, T.~M., \& {Sackett}, P.~D. 2010, \apj, 712, 38

\bibitem[{{Fortney} {et~al.}(2007){Fortney}, {Marley}, \& {Barnes}}]{fortney07}
{Fortney}, J.~J., {Marley}, M.~S., \& {Barnes}, J.~W. 2007, \apj, 659, 1661

\bibitem[{{Fressin} {et~al.}(2011){Fressin}, {Torres}, {Desert}, {Charbonneau},
  {Batalha}, {Fortney}, {Rowe}, {Allen}, {Borucki}, {Brown}, {Bryson},
  {Ciardi}, {Cochran}, {Deming}, {Dunham}, {Fabrycky}, {Gautier}, {Gilliland},
  {Henze}, {Holman}, {Howell}, {Jenkins}, {Kinemuchi}, {Knutson}, {Koch},
  {Latham}, {Lissauer}, {Marcy}, {Ragozzine}, {Sasselov}, {Still}, {Tenenbaum},
  \& {Uddin}}]{fressin11}
{Fressin}, F., {et~al.} 2011, ArXiv e-prints

\bibitem[{{Gaudi}(2005)}]{gaudi05}
{Gaudi}, B.~S. 2005, \apjl, 628, L73

\bibitem[{{Gaudi} {et~al.}(2005){Gaudi}, {Seager}, \&
  {Mallen-Ornelas}}]{gaudietal05}
{Gaudi}, B.~S., {Seager}, S., \& {Mallen-Ornelas}, G. 2005, \apj, 623, 472

\bibitem[{{Girardi} {et~al.}(2002){Girardi}, {Bertelli}, {Bressan}, {Chiosi},
  {Groenewegen}, {Marigo}, {Salasnich}, \& {Weiss}}]{girardi02}
{Girardi}, L., {Bertelli}, G., {Bressan}, A., {Chiosi}, C., {Groenewegen},
  M.~A.~T., {Marigo}, P., {Salasnich}, B., \& {Weiss}, A. 2002, \aap, 391, 195

\bibitem[{{Girardi} {et~al.}(2005){Girardi}, {Groenewegen}, {Hatziminaoglou},
  \& {da Costa}}]{girardi05}
{Girardi}, L., {Groenewegen}, M.~A.~T., {Hatziminaoglou}, E., \& {da Costa}, L.
  2005, \aap, 436, 895

\bibitem[{{Hartman} {et~al.}(2004){Hartman}, {Bakos}, {Stanek}, \&
  {Noyes}}]{hartman04}
{Hartman}, J.~D., {Bakos}, G., {Stanek}, K.~Z., \& {Noyes}, R.~W. 2004, \aj,
  128, 1761

\bibitem[{{Ho} \& {Turner}(2010)}]{hoturner}
{Ho}, S., \& {Turner}, E.~L. 2010, ArXiv e-prints

\bibitem[{{Howard} {et~al.}(2010){Howard}, {Marcy}, {Johnson}, {Fischer},
  {Wright}, {Isaacson}, {Valenti}, {Anderson}, {Lin}, \& {Ida}}]{howard10}
{Howard}, A.~W., {et~al.} 2010, Science, 330, 653

\bibitem[{{Jenkins} {et~al.}(2010{\natexlab{a}}){Jenkins}, {Borucki}, {Koch},
  {Marcy}, {Cochran}, {Welsh}, {Basri}, {Batalha}, {Buchhave}, {Brown},
  {Caldwell}, {Dunham}, {Endl}, {Fischer}, {Gautier}, {Geary}, {Gilliland},
  {Howell}, {Isaacson}, {Johnson}, {Latham}, {Lissauer}, {Monet}, {Rowe},
  {Sasselov}, {Howard}, {MacQueen}, {Orosz}, {Chandrasekaran}, {Twicken},
  {Bryson}, {Quintana}, {Clarke}, {Li}, {Allen}, {Tenenbaum}, {Wu}, {Meibom},
  {Klaus}, {Middour}, {Cote}, {McCauliff}, {Girouard}, {Gunter}, {Wohler},
  {Hall}, {Ibrahim}, {Kamal Uddin}, {Wu}, {Bhavsar}, {Van Cleve}, {Pletcher},
  {Dotson}, \& {Haas}}]{jenkins10}
{Jenkins}, J.~M., {et~al.} 2010{\natexlab{a}}, \apj, 724, 1108

\bibitem[{{Jenkins} {et~al.}(2010{\natexlab{b}}){Jenkins}, {Caldwell},
  {Chandrasekaran}, {Twicken}, {Bryson}, {Quintana}, {Clarke}, {Li}, {Allen},
  {Tenenbaum}, {Wu}, {Klaus}, {Van Cleve}, {Dotson}, {Haas}, {Gilliland},
  {Koch}, \& {Borucki}}]{jenkins10b}
---. 2010{\natexlab{b}}, \apjl, 713, L120

\bibitem[{{Johnson} {et~al.}(2010){Johnson}, {Aller}, {Howard}, \&
  {Crepp}}]{johnson10}
{Johnson}, J.~A., {Aller}, K.~M., {Howard}, A.~W., \& {Crepp}, J.~R. 2010,
  \pasp, 122, 905

\bibitem[{{Koch} {et~al.}(1998){Koch}, {Borucki}, {Webster}, {Dunham},
  {Jenkins}, {Marriott}, \& {Reitsema}}]{koch98}
{Koch}, D.~G., {Borucki}, W., {Webster}, L., {Dunham}, E., {Jenkins}, J.,
  {Marriott}, J., \& {Reitsema}, H.~J. 1998, in Presented at the Society of
  Photo-Optical Instrumentation Engineers (SPIE) Conference, Vol. 3356, Society
  of Photo-Optical Instrumentation Engineers (SPIE) Conference Series, ed.
  {P.~Y.~Bely \& J.~B.~Breckinridge}, 599--607

\bibitem[{{Konacki} {et~al.}(2003){Konacki}, {Torres}, {Sasselov}, \&
  {Jha}}]{OGLEFPs}
{Konacki}, M., {Torres}, G., {Sasselov}, D.~D., \& {Jha}, S. 2003, \apj, 597,
  1076

\bibitem[{{Latham} {et~al.}(2005){Latham}, {Brown}, {Monet}, {Everett},
  {Esquerdo}, \& {Hergenrother}}]{latham05}
{Latham}, D.~W., {Brown}, T.~M., {Monet}, D.~G., {Everett}, M., {Esquerdo},
  G.~A., \& {Hergenrother}, C.~W. 2005, in Bulletin of the American
  Astronomical Society, Vol.~37, Bulletin of the American Astronomical Society,
  1340--+

\bibitem[{{Latham} {et~al.}(2009){Latham}, {Bakos}, {Torres}, {Stefanik},
  {Noyes}, {Kov{\'a}cs}, {P{\'a}l}, {Marcy}, {Fischer}, {Butler}, {Sip{\H
  o}cz}, {Sasselov}, {Esquerdo}, {Vogt}, {Hartman}, {Kov{\'a}cs},
  {L{\'a}z{\'a}r}, {Papp}, \& {S{\'a}ri}}]{latham09}
{Latham}, D.~W., {et~al.} 2009, \apj, 704, 1107

\bibitem[{{Lissauer} {et~al.}(2011){Lissauer}, {Fabrycky}, {Ford}, {Borucki},
  {Fressin}, {Marcy}, {Orosz}, {Rowe}, {Torres}, {Welsh}, {Batalha}, {Bryson},
  {Buchhave}, {Caldwell}, {Carter}, {Charbonneau}, {Christiansen}, {Cochran},
  {Desert}, {Dunham}, {Fanelli}, {Fortney}, {Gautier}, {Geary}, {Gilliland},
  {Haas}, {Hall}, {Holman}, {Koch}, {Latham}, {Lopez}, {McCauliff}, {Miller},
  {Morehead}, {Quintana}, {Ragozzine}, {Sasselov}, {Short}, \&
  {Steffen}}]{lissauer11}
{Lissauer}, J.~J., {et~al.} 2011, \nat, 470, 53

\bibitem[{{Luhman} \& {Jayawardhana}(2002)}]{luhman02}
{Luhman}, K.~L., \& {Jayawardhana}, R. 2002, \apj, 566, 1132

\bibitem[{{O'Donovan} {et~al.}(2006){O'Donovan}, {Charbonneau}, {Torres},
  {Mandushev}, {Dunham}, {Latham}, {Alonso}, {Brown}, {Esquerdo}, {Everett}, \&
  {Creevey}}]{odonovan06}
{O'Donovan}, F.~T., {et~al.} 2006, \apj, 644, 1237

\bibitem[{{Poleski} {et~al.}(2010){Poleski}, {McCullough}, {Valenti}, {Burke},
  {Machalek}, \& {Janes}}]{XOFPs}
{Poleski}, R., {McCullough}, P.~R., {Valenti}, J.~A., {Burke}, C.~J.,
  {Machalek}, P., \& {Janes}, K. 2010, \apjs, 189, 134

\bibitem[{{Raghavan} {et~al.}(2010){Raghavan}, {McAlister}, {Henry}, {Latham},
  {Marcy}, {Mason}, {Gies}, {White}, \& {ten Brummelaar}}]{raghavan}
{Raghavan}, D., {et~al.} 2010, \apjs, 190, 1

\bibitem[{{Seager} {et~al.}(2007){Seager}, {Kuchner}, {Hier-Majumder}, \&
  {Militzer}}]{seager07}
{Seager}, S., {Kuchner}, M., {Hier-Majumder}, C.~A., \& {Militzer}, B. 2007,
  \apj, 669, 1279

\bibitem[{{Torres} {et~al.}(2004){Torres}, {Konacki}, {Sasselov}, \&
  {Jha}}]{torres04}
{Torres}, G., {Konacki}, M., {Sasselov}, D.~D., \& {Jha}, S. 2004, \apj, 614,
  979

\bibitem[{{Torres} {et~al.}(2011){Torres}, {Fressin}, {Batalha}, {Borucki},
  {Brown}, {Bryson}, {Buchhave}, {Charbonneau}, {Ciardi}, {Dunham}, {Fabrycky},
  {Ford}, {Gautier}, {Gilliland}, {Holman}, {Howell}, {Isaacson}, {Jenkins},
  {Koch}, {Latham}, {Lissauer}, {Marcy}, {Monet}, {Prsa}, {Quinn}, {Ragozzine},
  {Rowe}, {Sasselov}, {Steffen}, \& {Welsh}}]{torres11}
{Torres}, G., {et~al.} 2011, \apj, 727, 24

\end{thebibliography}

\end{document}